\documentclass[%
 reprint,
superscriptaddress,
nofootinbib,
 amsmath,amssymb,
 aps,
]{revtex4-2}
\usepackage{subcaption}
\usepackage{graphicx}
\usepackage{dcolumn}
\usepackage{bm}

\usepackage{color}
\usepackage{hyperref}
\usepackage{cleveref}
\usepackage{comment}
\usepackage[normalem]{ulem}
\renewcommand{\l}{\left}
\renewcommand{\r}{\right}

\newcommand{\xmark}{\text{\sffamily X}}%

\begin{document}

\newcommand{\chirpTFhpd}{$1.1869^{+0.0035}_{-0.0023}$}
\newcommand{\chirpTFsym}{$1.1869^{+0.0038}_{-0.0019}$}
\newcommand{\chirpTFsigYagihpd}{$1.1870^{+0.0035}_{-0.0023}$}
\newcommand{\chirpTFsigYagisym}{$1.1870^{+0.0037}_{-0.0020}$}
\newcommand{\chirpTFsigThishpd}{$1.1869^{+0.0034}_{-0.0023}$}
\newcommand{\chirpTFsigThissym}{$1.1869^{+0.0038}_{-0.0019}$}
\newcommand{\chirpTFfmtidalYagiChanhpd}{$1.1869^{+0.0034}_{-0.0023}$}
\newcommand{\chirpTFfmtidalYagiChansym}{$1.1869^{+0.0038}_{-0.0019}$}
\newcommand{\chirpTFfmtidalThishpd}{$1.1870^{+0.0035}_{-0.0024}$}
\newcommand{\chirpTFfmtidalThissym}{$1.1870^{+0.0038}_{-0.0020}$}
\newcommand{\qTFhpd}{(0.74,1.00)}
\newcommand{\qTFsym}{(0.72,1.00)}
\newcommand{\qTFsigYagihpd}{(0.73,1.00)}
\newcommand{\qTFsigYagisym}{(0.72,1.00)}
\newcommand{\qTFsigThishpd}{(0.72,1.00)}
\newcommand{\qTFsigThissym}{(0.71,1.00)}
\newcommand{\qTFfmtidalYagiChanhpd}{(0.72,1.00)}
\newcommand{\qTFfmtidalYagiChansym}{(0.71,1.00)}
\newcommand{\qTFfmtidalThishpd}{(0.74,1.00)}
\newcommand{\qTFfmtidalThissym}{(0.72,1.00)}
\newcommand{\chieffTFhpd}{$0.00^{+0.01}_{-0.01}$}
\newcommand{\chieffTFsym}{$0.00^{+0.01}_{-0.01}$}
\newcommand{\chieffTFsigYagihpd}{$0.00^{+0.01}_{-0.01}$}
\newcommand{\chieffTFsigYagisym}{$0.00^{+0.01}_{-0.01}$}
\newcommand{\chieffTFsigThishpd}{$0.00^{+0.01}_{-0.01}$}
\newcommand{\chieffTFsigThissym}{$0.00^{+0.01}_{-0.01}$}
\newcommand{\chieffTFfmtidalYagiChanhpd}{$0.00^{+0.01}_{-0.01}$}
\newcommand{\chieffTFfmtidalYagiChansym}{$0.00^{+0.01}_{-0.01}$}
\newcommand{\chieffTFfmtidalThishpd}{$0.00^{+0.01}_{-0.01}$}
\newcommand{\chieffTFfmtidalThissym}{$0.00^{+0.01}_{-0.01}$}
\newcommand{\thetajnTFhpd}{$2.55^{+0.47}_{-0.46}$}
\newcommand{\thetajnTFsym}{$2.55^{+0.42}_{-0.46}$}
\newcommand{\thetajnTFsigYagihpd}{$2.54^{+0.45}_{-0.47}$}
\newcommand{\thetajnTFsigYagisym}{$2.54^{+0.42}_{-0.45}$}
\newcommand{\thetajnTFsigThishpd}{$2.56^{+0.46}_{-0.44}$}
\newcommand{\thetajnTFsigThissym}{$2.56^{+0.42}_{-0.46}$}
\newcommand{\thetajnTFfmtidalYagiChanhpd}{$2.55^{+0.48}_{-0.47}$}
\newcommand{\thetajnTFfmtidalYagiChansym}{$2.55^{+0.43}_{-0.45}$}
\newcommand{\thetajnTFfmtidalThishpd}{$2.54^{+0.47}_{-0.46}$}
\newcommand{\thetajnTFfmtidalThissym}{$2.54^{+0.43}_{-0.46}$}
\newcommand{\dlTFhpd}{$6^{+245}_{-235}$}
\newcommand{\dlTFsym}{$6^{+242}_{-225}$}
\newcommand{\dlTFsigYagihpd}{$7^{+256}_{-212}$}
\newcommand{\dlTFsigYagisym}{$7^{+226}_{-216}$}
\newcommand{\dlTFsigThishpd}{$7^{+240}_{-223}$}
\newcommand{\dlTFsigThissym}{$7^{+231}_{-221}$}
\newcommand{\dlTFfmtidalYagiChanhpd}{$8^{+208}_{-193}$}
\newcommand{\dlTFfmtidalYagiChansym}{$8^{+201}_{-185}$}
\newcommand{\dlTFfmtidalThishpd}{$5^{+186}_{-197}$}
\newcommand{\dlTFfmtidalThissym}{$5^{+198}_{-184}$}
\newcommand{\ldTFhpd}{$40^{+8}_{-12}$}
\newcommand{\ldTFsym}{$40^{+7}_{-14}$}
\newcommand{\ldTFsigYagihpd}{$39^{+8}_{-12}$}
\newcommand{\ldTFsigYagisym}{$39^{+7}_{-14}$}
\newcommand{\ldTFsigThishpd}{$40^{+8.54}_{-13.49}$}
\newcommand{\ldTFsigThissym}{$40^{+7}_{-14}$}
\newcommand{\ldTFfmtidalYagiChanhpd}{$39^{+8}_{-12.09}$}
\newcommand{\ldTFfmtidalYagiChansym}{$39^{+7}_{-14}$}
\newcommand{\ldTFfmtidalThishpd}{$39.69^{+8.89}_{-13.11}$}
\newcommand{\ldTFfmtidalThissym}{$39^{+7}_{-14}$}
\newcommand{\LtildeTFhpd}{$422^{+557}_{-296}$}
\newcommand{\LtildeTFsym}{$422^{+659}_{-247}$}
\newcommand{\LtildeTFsigYagihpd}{$419^{+522}_{-282}$}
\newcommand{\LtildeTFsigYagisym}{$419^{+645}_{-246}$}
\newcommand{\LtildeTFsigThishpd}{$418^{+572}_{-295}$}
\newcommand{\LtildeTFsigThissym}{$418^{+638}_{-253}$}
\newcommand{\LtildeTFfmtidalYagiChanhpd}{$377^{+389}_{-267}$}
\newcommand{\LtildeTFfmtidalYagiChansym}{$377^{+475}_{-244}$}
\newcommand{\LtildeTFfmtidalThishpd}{$400^{+417}_{-286}$}
\newcommand{\LtildeTFfmtidalThissym}{$400^{+450}_{-267}$}

\newcommand{\RprimeYagiMasseli}{$11.99^{+2.84}_{-3.22}$}
\newcommand{\RprimeYagiMasselisym}{$11.99^{+2.59}_{-3.14}$}
\newcommand{\RprimeTFthis}{$12.19^{+3.48}_{-3.12}$}
\newcommand{\RprimeTFthissym}{$12.19^{+2.97}_{-3.25}$}
\newcommand{\RprimeTFMaselli}{$12.00^{+2.91}_{-3.19}$}
\newcommand{\RprimeTFMasellisym}{$12.00^{+2.65}_{-3.14}$}
\newcommand{\Rprimesigthiswork}{$12.19^{+3.14}_{-3.37}$}
\newcommand{\Rprimesigthisworksym}{$12.19^{+2.99}_{-3.29}$}
\newcommand{\RprimefmYagiChan}{$11.59^{+2.83}_{-2.99}$}
\newcommand{\RprimefmYagiChansym}{$11.59^{+2.49}_{-3.12}$}
\newcommand{\Rprimefmthiswork}{$11.89^{+2.85}_{-2.83}$}
\newcommand{\Rprimefmthisworksym}{$11.89^{+2.61}_{-3.14}$}
\newcommand{\RsecondYagiMasseli}{$11.07^{+2.98}_{-2.78}$}
\newcommand{\RsecondYagiMasselisym}{$11.07^{+2.64}_{-3.05}$}
\newcommand{\RsecondTFMaselli}{$11.08^{+3.15}_{-3.01}$}
\newcommand{\RsecondTFMasellisym}{$11.08^{+2.74}_{-3.07}$}
\newcommand{\RsecondTFthis}{$11.32^{+3.58}_{-3.18}$}
\newcommand{\RsecondTFthissym}{$11.32^{+3.08}_{-3.25}$}
\newcommand{\Rsecondsigthiswork}{$11.29^{+3.43}_{-3.13}$}
\newcommand{\Rsecondsigthisworksym}{$11.29^{+3.01}_{-3.30}$}
\newcommand{\RsecondfmYagiChan}{$10.67^{+2.67}_{-2.79}$}
\newcommand{\RsecondfmYagiChansym}{$10.67^{+2.51}_{-2.95}$}
\newcommand{\Rsecondfmthiswork}{$10.91^{+2.98}_{-3.35}$}
\newcommand{\Rsecondfmthisworksym}{$10.91^{+2.71}_{-3.30}$}

\newcommand{\chirpfmnewmconezerosixAPR}{$1.067^{+0.0001}_{-0.0001}$}
\newcommand{\qfmnewmconezerosixAPR}{$0.856^{+0.013}_{-0.012}$}
\newcommand{\dlfmnewmconezerosixAPR}{$14^{+208}_{-207}$}
\newcommand{\chirpfmnewmconeeightAPR}{$1.186^{+0.0001}_{-0.0001}$}
\newcommand{\qfmnewmconeeightAPR}{$0.855^{+0.013}_{-0.012}$}
\newcommand{\dlfmnewmconeeightAPR}{$8^{+110}_{-113}$}
\newcommand{\chirpfmnewmconethreeAPR}{$1.304^{+0.0001}_{-0.0001}$}
\newcommand{\qfmnewmconethreeAPR}{$0.855^{+0.012}_{-0.011}$}
\newcommand{\dlfmnewmconethreeAPR}{$1^{+61}_{-59}$}
\newcommand{\chirpfmoldmconezerosixAPR}{$1.067^{+0.0001}_{-0.0001}$}
\newcommand{\chirpinjectmconezerosixAPR}{$1.067$}
\newcommand{\qfmoldmconezerosixAPR}{$0.856^{+0.013}_{-0.012}$}
\newcommand{\qinjectmconezerosixAPR}{$0.855$}
\newcommand{\dlfmoldmconezerosixAPR}{$13^{+204}_{-208}$}
\newcommand{\dlinjectmconezerosixAPR}{$62.439$}
\newcommand{\chirpfmoldmconeeightAPR}{$1.186^{+0.0001}_{-0.0001}$}
\newcommand{\chirpinjectmconeeightAPR}{$1.186$}
\newcommand{\qfmoldmconeeightAPR}{$0.855^{+0.013}_{-0.011}$}
\newcommand{\qinjectmconeeightAPR}{$0.855$}
\newcommand{\dlfmoldmconeeightAPR}{$5^{+113}_{-111}$}
\newcommand{\dlinjectmconeeightAPR}{$37.390$}
\newcommand{\chirpfmoldmconethreeAPR}{$1.304^{+0.0001}_{-0.0001}$}
\newcommand{\chirpinjectmconethreeAPR}{$1.304$}
\newcommand{\qfmoldmconethreeAPR}{$0.854^{+0.012}_{-0.011}$}
\newcommand{\qinjectmconethreeAPR}{$0.855$}
\newcommand{\dlfmoldmconethreeAPR}{$3^{+61}_{-61}$}
\newcommand{\dlinjectmconethreeAPR}{$23.114$}
\newcommand{\chirpnofmmconezerosixAPR}{$1.067^{+0.0001}_{-0.0001}$}
\newcommand{\qnofmmconezerosixAPR}{$0.848^{+0.012}_{-0.012}$}
\newcommand{\dlnofmmconezerosixAPR}{$-2^{+258}_{-232}$}
\newcommand{\chirpnofmmconeeightAPR}{$1.186^{+0.0001}_{-0.0001}$}
\newcommand{\qnofmmconeeightAPR}{$0.848^{+0.013}_{-0.011}$}
\newcommand{\dlnofmmconeeightAPR}{$5^{+131}_{-127}$}
\newcommand{\chirpnofmmconethreeAPR}{$1.304^{+0.0001}_{-0.0001}$}
\newcommand{\qnofmmconethreeAPR}{$0.851^{+0.012}_{-0.011}$}
\newcommand{\dlnofmmconethreeAPR}{$2^{+69}_{-66}$}
\newcommand{\LfmnewmconezerosixAPR}{$581^{+27}_{-28}$}
\newcommand{\LfmnewmconeeightAPR}{$307^{+22}_{-22}$}
\newcommand{\LfmnewmconethreeAPR}{$169^{+18}_{-18}$}
\newcommand{\LfmoldmconezerosixAPR}{$587^{+28}_{-30}$}
\newcommand{\LinjectmconezerosixAPR}{$590.35$}
\newcommand{\LfmoldmconeeightAPR}{$311^{+21}_{-22}$}
\newcommand{\LinjectmconeeightAPR}{$313.94$}
\newcommand{\LfmoldmconethreeAPR}{$165^{+18}_{-19}$}
\newcommand{\LinjectmconethreeAPR}{$168.01$}
\newcommand{\LnofmmconezerosixAPR}{$684^{+32}_{-34}$}
\newcommand{\LnofmmconeeightAPR}{$358^{+24}_{-25}$}
\newcommand{\LnofmmconethreeAPR}{$184^{+21}_{-20}$}

\newcommand{\chirpfmnewmconezerosixStiff}{$1.067^{+0.0001}_{-0.0001}$}
\newcommand{\qfmnewmconezerosixStiff}{$0.860^{+0.016}_{-0.013}$}
\newcommand{\dlfmnewmconezerosixStiff}{$69^{+469}_{-495}$}
\newcommand{\chirpfmnewmconeeightStiff}{$1.186^{+0.0001}_{-0.0001}$}
\newcommand{\qfmnewmconeeightStiff}{$0.858^{+0.014}_{-0.012}$}
\newcommand{\dlfmnewmconeeightStiff}{$50^{+291}_{-308}$}
\newcommand{\chirpfmnewmconethreeStiff}{$1.304^{+0.0001}_{-0.0001}$}
\newcommand{\qfmnewmconethreeStiff}{$0.863^{+0.014}_{-0.013}$}
\newcommand{\dlfmnewmconethreeStiff}{$8^{+210}_{-177}$}
\newcommand{\chirpfmoldmconezerosixStiff}{$1.067^{+0.0001}_{-0.0001}$}
\newcommand{\chirpinjectmconezerosixStiff}{$1.067$}
\newcommand{\qfmoldmconezerosixStiff}{$0.859^{+0.015}_{-0.013}$}
\newcommand{\qinjectmconezerosixStiff}{$0.855$}
\newcommand{\dlfmoldmconezerosixStiff}{$67^{+500}_{-498}$}
\newcommand{\dlinjectmconezerosixStiff}{$166.750$}
\newcommand{\chirpfmoldmconeeightStiff}{$1.186^{+0.0001}_{-0.0001}$}
\newcommand{\chirpinjectmconeeightStiff}{$1.186$}
\newcommand{\qfmoldmconeeightStiff}{$0.858^{+0.014}_{-0.013}$}
\newcommand{\qinjectmconeeightStiff}{$0.855$}
\newcommand{\dlfmoldmconeeightStiff}{$41^{+308}_{-313}$}
\newcommand{\dlinjectmconeeightStiff}{$83.325$}
\newcommand{\chirpfmoldmconethreeStiff}{$1.304^{+0.0001}_{-0.0001}$}
\newcommand{\chirpinjectmconethreeStiff}{$1.304$}
\newcommand{\qfmoldmconethreeStiff}{$0.853^{+0.012}_{-0.010}$}
\newcommand{\qinjectmconethreeStiff}{$0.855$}
\newcommand{\dlfmoldmconethreeStiff}{$54^{+193}_{-224}$}
\newcommand{\dlinjectmconethreeStiff}{$65.100$}
\newcommand{\chirpnofmmconezerosixStiff}{$1.067^{+0.0001}_{-0.0001}$}
\newcommand{\qnofmmconezerosixStiff}{$0.830^{+0.013}_{-0.012}$}
\newcommand{\dlnofmmconezerosixStiff}{$-65^{+599}_{-639}$}
\newcommand{\chirpnofmmconeeightStiff}{$1.186^{+0.0001}_{-0.0001}$}
\newcommand{\qnofmmconeeightStiff}{$0.833^{+0.012}_{-0.011}$}
\newcommand{\dlnofmmconeeightStiff}{$-10^{+373}_{-355}$}
\newcommand{\chirpnofmmconethreeStiff}{$1.304^{+0.0001}_{-0.0001}$}
\newcommand{\qnofmmconethreeStiff}{$0.842^{+0.012}_{-0.011}$}
\newcommand{\dlnofmmconethreeStiff}{$-1^{+166}_{-159}$}
\newcommand{\LfmnewmconezerosixStiff}{$1983^{+33}_{-45}$}
\newcommand{\LfmnewmconeeightStiff}{$1076^{+24}_{-31}$}
\newcommand{\LfmnewmconethreeStiff}{$713^{+21}_{-22}$}
\newcommand{\LfmoldmconezerosixStiff}{$1975^{+33}_{-44}$}
\newcommand{\LinjectmconezerosixStiff}{$2005.46$}
\newcommand{\LfmoldmconeeightStiff}{$1069^{+25}_{-30}$}
\newcommand{\LinjectmconeeightStiff}{$1087.65$}
\newcommand{\LfmoldmconethreeStiff}{$721^{+19}_{-27}$}
\newcommand{\LinjectmconethreeStiff}{$720.18$}
\newcommand{\LnofmmconezerosixStiff}{$2289^{+51}_{-56}$}
\newcommand{\LnofmmconeeightStiff}{$1258^{+31}_{-34}$}
\newcommand{\LnofmmconethreeStiff}{$811^{+23}_{-25}$}

\newcommand{\chirpfmnewmconezerosixSoft}{$1.067^{+0.0001}_{-0.0001}$}
\newcommand{\qfmnewmconezerosixSoft}{$0.859^{+0.014}_{-0.012}$}
\newcommand{\dlfmnewmconezerosixSoft}{$7^{+89}_{-92}$}
\newcommand{\chirpfmnewmconeeightSoft}{$1.186^{+0.0001}_{-0.0001}$}
\newcommand{\qfmnewmconeeightSoft}{$0.846^{+0.010}_{-0.009}$}
\newcommand{\dlfmnewmconeeightSoft}{$4^{+54}_{-55}$}
\newcommand{\chirpfmnewmconethreeSoft}{$1.304^{+0.0001}_{-0.0001}$}
\newcommand{\qfmnewmconethreeSoft}{$0.857^{+0.012}_{-0.011}$}
\newcommand{\dlfmnewmconethreeSoft}{$1^{+29}_{-28}$}
\newcommand{\chirpfmoldmconezerosixSoft}{$1.067^{+0.0001}_{-0.0001}$}
\newcommand{\chirpinjectmconezerosixSoft}{$1.067$}
\newcommand{\qfmoldmconezerosixSoft}{$0.861^{+0.014}_{-0.012}$}
\newcommand{\qinjectmconezerosixSoft}{$0.855$}
\newcommand{\dlfmoldmconezerosixSoft}{$6^{+92}_{-92}$}
\newcommand{\dlinjectmconezerosixSoft}{$24.613$}
\newcommand{\chirpfmoldmconeeightSoft}{$1.186^{+0.0001}_{-0.0001}$}
\newcommand{\chirpinjectmconeeightSoft}{$1.186$}
\newcommand{\qfmoldmconeeightSoft}{$0.850^{+0.011}_{-0.009}$}
\newcommand{\qinjectmconeeightSoft}{$0.855$}
\newcommand{\dlfmoldmconeeightSoft}{$6^{+51}_{-55}$}
\newcommand{\dlinjectmconeeightSoft}{$15.428$}
\newcommand{\chirpfmoldmconethreeSoft}{$1.304^{+0.0001}_{-0.0001}$}
\newcommand{\chirpinjectmconethreeSoft}{$1.304$}
\newcommand{\qfmoldmconethreeSoft}{$0.846^{+0.011}_{-0.011}$}
\newcommand{\qinjectmconethreeSoft}{$0.854$}
\newcommand{\dlfmoldmconethreeSoft}{$2^{+30}_{-30}$}
\newcommand{\dlinjectmconethreeSoft}{$9.622$}
\newcommand{\chirpnofmmconezerosixSoft}{$1.067^{+0.0001}_{-0.0001}$}
\newcommand{\qnofmmconezerosixSoft}{$0.839^{+0.011}_{-0.010}$}
\newcommand{\dlnofmmconezerosixSoft}{$5^{+120}_{-115}$}
\newcommand{\chirpnofmmconeeightSoft}{$1.186^{+0.0001}_{-0.0001}$}
\newcommand{\qnofmmconeeightSoft}{$0.854^{+0.011}_{-0.010}$}
\newcommand{\dlnofmmconeeightSoft}{$3^{+57}_{-56}$}
\newcommand{\chirpnofmmconethreeSoft}{$1.304^{+0.0001}_{-0.0001}$}
\newcommand{\qnofmmconethreeSoft}{$0.849^{+0.012}_{-0.011}$}
\newcommand{\dlnofmmconethreeSoft}{$2^{+37}_{-36}$}
\newcommand{\LfmnewmconezerosixSoft}{$257^{+29}_{-29}$}
\newcommand{\LfmnewmconeeightSoft}{$151^{+18}_{-19}$}
\newcommand{\LfmnewmconethreeSoft}{$79^{+18}_{-18}$}
\newcommand{\LfmoldmconezerosixSoft}{$249^{+23}_{-23}$}
\newcommand{\LinjectmconezerosixSoft}{$258.95$}
\newcommand{\LfmoldmconeeightSoft}{$145^{+19}_{-18}$}
\newcommand{\LinjectmconeeightSoft}{$144.84$}
\newcommand{\LfmoldmconethreeSoft}{$82^{+19}_{-20}$}
\newcommand{\LinjectmconethreeSoft}{$80.91$}
\newcommand{\LnofmmconezerosixSoft}{$317^{+25}_{-27}$}
\newcommand{\LnofmmconeeightSoft}{$159^{+23}_{-25}$}
\newcommand{\LnofmmconethreeSoft}{$98^{+24}_{-23}$}

\newcommand{\chirpfmnewmconezerosixETAPR}{$1.067^{+\rm 1E-5}_{-\rm 1E-5}$}
\newcommand{\qfmnewmconezerosixETAPR}{$0.856^{+0.005}_{-0.005}$}
\newcommand{\dlfmnewmconezerosixETAPR}{$28^{+177}_{-188}$}
\newcommand{\chirpfmoldmconezerosixETAPR}{$1.067^{+\rm 1E-5}_{-\rm 1E-5}$}
\newcommand{\chirpinjectmconezerosixETAPR}{$1.067$}
\newcommand{\qfmoldmconezerosixETAPR}{$0.856^{+0.005}_{-0.005}$}
\newcommand{\qinjectmconezerosixETAPR}{$0.855$}
\newcommand{\dlfmoldmconezerosixETAPR}{$30^{+176}_{-183}$}
\newcommand{\dLinjectmconezerosixETAPR}{$62.420$}
\newcommand{\chirpnofmmconezerosixETAPR}{$1.067^{+\rm 1E-5}_{-\rm 1E-5}$}
\newcommand{\qnofmmconezerosixETAPR}{$0.843^{+0.005}_{-0.004}$}
\newcommand{\dlnofmmconezerosixETAPR}{$-37^{+213}_{-204}$}
\newcommand{\chirpfmnewmconeeightETAPR}{$1.186^{+\rm 1E-5}_{-\rm 1E-5}$}
\newcommand{\qfmnewmconeeightETAPR}{$0.855^{+0.005}_{-0.004}$}
\newcommand{\dlfmnewmconeeightETAPR}{$13^{+106}_{-112}$}
\newcommand{\chirpfmoldmconeeightETAPR}{$1.186^{+\rm 1E-5}_{-\rm 1E-5}$}
\newcommand{\chirpinjectmconeeightETAPR}{$1.186$}
\newcommand{\qfmoldmconeeightETAPR}{$0.855^{+0.005}_{-0.004}$}
\newcommand{\qinjectmconeeightETAPR}{$0.855$}
\newcommand{\dlfmoldmconeeightETAPR}{$10^{+107}_{-108}$}
\newcommand{\dLinjectmconeeightETAPR}{$31.189$}
\newcommand{\chirpnofmmconeeightETAPR}{$1.186^{+\rm 1E-5}_{-\rm 1E-5}$}
\newcommand{\qnofmmconeeightETAPR}{$0.848^{+0.005}_{-0.004}$}
\newcommand{\dlnofmmconeeightETAPR}{$-16^{+148}_{-111}$}
\newcommand{\chirpfmnewmconethreeETAPR}{$1.304^{+\rm 1E-5}_{-\rm 1E-5}$}
\newcommand{\qfmnewmconethreeETAPR}{$0.855^{+0.005}_{-0.004}$}
\newcommand{\dlfmnewmconethreeETAPR}{$5^{+59}_{-61}$}
\newcommand{\chirpfmoldmconethreeETAPR}{$1.304^{+\rm 1E-5}_{-\rm 1E-5}$}
\newcommand{\chirpinjectmconethreeETAPR}{$1.304$}
\newcommand{\qfmoldmconethreeETAPR}{$0.855^{+0.005}_{-0.004}$}
\newcommand{\qinjectmconethreeETAPR}{$0.854$}
\newcommand{\dlfmoldmconethreeETAPR}{$7^{+57}_{-64}$}
\newcommand{\dLinjectmconethreeETAPR}{$23.092$}
\newcommand{\chirpnofmmconethreeETAPR}{$1.304^{+\rm 1E-5}_{-\rm 1E-5}$}
\newcommand{\qnofmmconethreeETAPR}{$0.851^{+0.004}_{-0.004}$}
\newcommand{\dlnofmmconethreeETAPR}{$1^{+70}_{-65}$}
\newcommand{\LfmnewmconezerosixETAPR}{$588^{+11}_{-15}$}
\newcommand{\LfmoldmconezerosixETAPR}{$584^{+10}_{-15}$}
\newcommand{\LinjectmconezerosixETAPR}{$590$}
\newcommand{\LnofmmconezerosixETAPR}{$685^{+11}_{-12}$}
\newcommand{\LfmnewmconeeightETAPR}{$310^{+8}_{-10}$}
\newcommand{\LfmoldmconeeightETAPR}{$315^{+8}_{-9}$}
\newcommand{\LinjectmconeeightETAPR}{$314$}
\newcommand{\LnofmmconeeightETAPR}{$360^{+9}_{-8}$}
\newcommand{\LfmnewmconethreeETAPR}{$168^{+7}_{-7}$}
\newcommand{\LfmoldmconethreeETAPR}{$164^{+7}_{-7}$}
\newcommand{\LinjectmconethreeETAPR}{$168$}
\newcommand{\LnofmmconethreeETAPR}{$186^{+7}_{-7}$}
\newcommand{\chirpfmnewmconezerosixETSoft}{$1.067^{+\rm 1E-5}_{-\rm 1E-5}$}
\newcommand{\qfmnewmconezerosixETSoft}{$0.856^{+0.005}_{-0.005}$}
\newcommand{\dlfmnewmconezerosixETSoft}{$4^{+92}_{-91}$}
\newcommand{\chirpfmoldmconezerosixETSoft}{$1.067^{+\rm 1E-5}_{-\rm 1E-5}$}
\newcommand{\chirpinjectmconezerosixETSoft}{$1.067$}
\newcommand{\qfmoldmconezerosixETSoft}{$0.856^{+0.005}_{-0.005}$}
\newcommand{\qinjectmconezerosixETSoft}{$0.855$}
\newcommand{\dlfmoldmconezerosixETSoft}{$5^{+94}_{-89}$}
\newcommand{\dLinjectmconezerosixETSoft}{$24.604$}
\newcommand{\chirpnofmmconezerosixETSoft}{$1.067^{+\rm 1E-5}_{-\rm 1E-5}$}
\newcommand{\qnofmmconezerosixETSoft}{$0.851^{+0.005}_{-0.005}$}
\newcommand{\dlnofmmconezerosixETSoft}{$-1^{+111}_{-101}$}
\newcommand{\chirpfmnewmconeeightETSoft}{$1.186^{+\rm 1E-5}_{-\rm 1E-5}$}
\newcommand{\qfmnewmconeeightETSoft}{$0.855^{+0.005}_{-0.005}$}
\newcommand{\dlfmnewmconeeightETSoft}{$3^{+51}_{-53}$}
\newcommand{\chirpfmoldmconeeightETSoft}{$1.186^{+\rm 1E-5}_{-\rm 1E-5}$}
\newcommand{\chirpinjectmconeeightETSoft}{$1.186$}
\newcommand{\qfmoldmconeeightETSoft}{$0.855^{+0.005}_{-0.004}$}
\newcommand{\qinjectmconeeightETSoft}{$0.855$}
\newcommand{\dlfmoldmconeeightETSoft}{$5^{+51}_{-53}$}
\newcommand{\dLinjectmconeeightETSoft}{$15.426$}
\newcommand{\chirpnofmmconeeightETSoft}{$1.186^{+\rm 1E-5}_{-\rm 1E-5}$}
\newcommand{\qnofmmconeeightETSoft}{$0.853^{+0.004}_{-0.005}$}
\newcommand{\dlnofmmconeeightETSoft}{$0.0^{+61}_{-56}$}
\newcommand{\chirpfmnewmconethreeETSoft}{$1.304^{+\rm 1E-5}_{-\rm 1E-5}$}
\newcommand{\qfmnewmconethreeETSoft}{$0.855^{+0.005}_{-0.004}$}
\newcommand{\dlfmnewmconethreeETSoft}{$2^{+28}_{-29}$}
\newcommand{\chirpfmoldmconethreeETSoft}{$1.304^{+\rm 1E-5}_{-\rm 1E-5}$}
\newcommand{\chirpinjectmconethreeETSoft}{$1.304$}
\newcommand{\qfmoldmconethreeETSoft}{$0.854^{+0.004}_{-0.004}$}
\newcommand{\qinjectmconethreeETSoft}{$0.854$}
\newcommand{\dlfmoldmconethreeETSoft}{$1^{+29}_{-28}$}
\newcommand{\dLinjectmconethreeETSoft}{$9.620$}
\newcommand{\chirpnofmmconethreeETSoft}{$1.304^{+\rm 1E-5}_{-\rm 1E-5}$}
\newcommand{\qnofmmconethreeETSoft}{$0.853^{+0.005}_{-0.004}$}
\newcommand{\dlnofmmconethreeETSoft}{$2^{+31}_{-32}$}
\newcommand{\LfmnewmconezerosixETSoft}{$256^{+10}_{-10}$}
\newcommand{\LfmoldmconezerosixETSoft}{$253^{+10}_{-10}$}
\newcommand{\LinjectmconezerosixETSoft}{$259$}
\newcommand{\LnofmmconezerosixETSoft}{$289^{+10}_{-10}$}
\newcommand{\LfmnewmconeeightETSoft}{$144^{+8}_{-8}$}
\newcommand{\LfmoldmconeeightETSoft}{$141^{+7}_{-8}$}
\newcommand{\LinjectmconeeightETSoft}{$145$}
\newcommand{\LnofmmconeeightETSoft}{$161^{+8}_{-8}$}
\newcommand{\LfmnewmconethreeETSoft}{$82^{+7}_{-7}$}
\newcommand{\LfmoldmconethreeETSoft}{$79^{+6}_{-6}$}
\newcommand{\LinjectmconethreeETSoft}{$81$}
\newcommand{\LnofmmconethreeETSoft}{$92^{+7}_{-7}$}

\newcommand{\chirpfmnewmconezerosixETStiff}{$1.067^{+\rm 1E-5}_{-\rm 1E-5}$}
\newcommand{\qfmnewmconezerosixETStiff}{$0.859^{+0.005}_{-0.005}$}
\newcommand{\dlfmnewmconezerosixETStiff}{$95^{+285}_{-303}$}
\newcommand{\chirpfmoldmconezerosixETStiff}{$1.067^{+\rm 1E-5}_{-\rm 1E-5}$}
\newcommand{\chirpinjectmconezerosixETStiff}{$1.067$}
\newcommand{\qfmoldmconezerosixETStiff}{$0.859^{+0.006}_{-0.005}$}
\newcommand{\qinjectmconezerosixETStiff}{$0.855$}
\newcommand{\dlfmoldmconezerosixETStiff}{$100^{+298}_{-298}$}
\newcommand{\dLinjectmconezerosixETStiff}{$166.686$}
\newcommand{\chirpnofmmconezerosixETStiff}{$1.067^{+\rm 1E-5}_{-\rm 1E-5}$}
\newcommand{\qnofmmconezerosixETStiff}{$0.826^{+0.005}_{-0.004}$}
\newcommand{\dlnofmmconezerosixETStiff}{$-34^{+188}_{-629}$}
\newcommand{\chirpfmnewmconeeightETStiff}{$1.186^{+\rm 1E-5}_{-\rm 1E-5}$}
\newcommand{\qfmnewmconeeightETStiff}{$0.856^{+0.005}_{-0.005}$}
\newcommand{\dlfmnewmconeeightETStiff}{$63^{+222}_{-241}$}
\newcommand{\chirpfmoldmconeeightETStiff}{$1.186^{+\rm 1E-5}_{-\rm 1E-5}$}
\newcommand{\chirpinjectmconeeightETStiff}{$1.186$}
\newcommand{\qfmoldmconeeightETStiff}{$0.857^{+0.005}_{-0.005}$}
\newcommand{\qinjectmconeeightETStiff}{$0.855$}
\newcommand{\dlfmoldmconeeightETStiff}{$61^{+224}_{-236}$}
\newcommand{\dLinjectmconeeightETStiff}{$82.886$}
\newcommand{\chirpnofmmconeeightETStiff}{$1.186^{+\rm 1E-5}_{-\rm 1E-5}$}
\newcommand{\qnofmmconeeightETStiff}{$0.831^{+0.004}_{-0.004}$}
\newcommand{\dlnofmmconeeightETStiff}{$-20^{+128}_{-157}$}
\newcommand{\chirpfmnewmconethreeETStiff}{$1.304^{+\rm 1E-5}_{-\rm 1E-5}$}
\newcommand{\qfmnewmconethreeETStiff}{$0.857^{+0.005}_{-0.004}$}
\newcommand{\dlfmnewmconethreeETStiff}{$47^{+146}_{-164}$}
\newcommand{\chirpfmoldmconethreeETStiff}{$1.304^{+\rm 1E-5}_{-\rm 1E-5}$}
\newcommand{\chirpinjectmconethreeETStiff}{$1.304$}
\newcommand{\qfmoldmconethreeETStiff}{$0.857^{+0.005}_{-0.005}$}
\newcommand{\qinjectmconethreeETStiff}{$0.854$}
\newcommand{\dlfmoldmconethreeETStiff}{$44^{+148}_{-162}$}
\newcommand{\dLinjectmconethreeETStiff}{$65.087$}
\newcommand{\chirpnofmmconethreeETStiff}{$1.304^{+\rm 1E-5}_{-\rm 1E-5}$}
\newcommand{\qnofmmconethreeETStiff}{$0.835^{+0.004}_{-0.004}$}
\newcommand{\dlnofmmconethreeETStiff}{$-28^{+119}_{-249}$}
\newcommand{\LfmnewmconezerosixETStiff}{$1993^{+13}_{-15}$}
\newcommand{\LfmoldmconezerosixETStiff}{$1985^{+11}_{-16}$}
\newcommand{\LinjectmconezerosixETStiff}{$2005$}
\newcommand{\LnofmmconezerosixETStiff}{$2285^{+14}_{-20}$}
\newcommand{\LfmnewmconeeightETStiff}{$1080^{+9}_{-15}$}
\newcommand{\LfmoldmconeeightETStiff}{$1075^{+9}_{-14}$}
\newcommand{\LinjectmconeeightETStiff}{$1087$}
\newcommand{\LnofmmconeeightETStiff}{$1261^{+9}_{-10}$}
\newcommand{\LfmnewmconethreeETStiff}{$715^{+7}_{-10}$}
\newcommand{\LfmoldmconethreeETStiff}{$712^{+8}_{-10}$}
\newcommand{\LinjectmconethreeETStiff}{$720$}
\newcommand{\LnofmmconethreeETStiff}{$825^{+7}_{-8}$}


\title{Impact of updated Multipole Love numbers and f-Love Universal Relations in the context of Binary Neutron Stars}

\author{Bikram Keshari Pradhan}
\email{bikramp@iucaa.in}
\affiliation{Inter-University Centre for Astronomy and Astrophysics,Pune University Campus,Pune,411007, India}
\author{Aditya Vijaykumar}
\affiliation{International Centre for Theoretical Sciences, Tata Institute of Fundamental Research, Bangalore 560089, India}
\affiliation{Department of Physics, The University of Chicago, 5640 South Ellis Avenue, Chicago, Illinois 60637, USA}
\author{Debarati Chatterjee}
\affiliation{Inter-University Centre for Astronomy and Astrophysics,Pune University Campus,Pune,411007, India}

%
%

\date{\today}

\begin{abstract}
Neutron star (NS) equation of state (EoS) insensitive relations or universal relations (UR) involving  neutron star bulk properties play a crucial role in gravitational-wave astronomy. Considering a wide range of equations of state originating from (i) phenomenological relativistic mean field models, (ii) realistic EoS models based on different physical motivations, and (iii) polytropic EoSs described by spectral decomposition method, we update the EoS-insensitive relations involving NS tidal deformability (Multipole Love relation) and the UR between f-mode frequency and tidal deformability (f-Love relation). We analyze the binary neutron star (BNS) event GW170817  using  the frequency domain TaylorF2 waveform model with updated universal relations and find that the additional contribution of the octupolar electric tidal parameter and quadrupolar magnetic tidal parameter or the change of multipole Love relation  has no significant impact on the inferred NS properties. However, adding the f-mode dynamical phase lowers the 90\% upper bound on $\tilde{\Lambda}$ by 16-20\% as well as lowers the upper bound of NSs radii by $\sim$500m. The combined URs (multipole Love and f-Love) developed in this work predict a higher median (also a higher 90\% upper bound)  for $\tilde{\Lambda}$  by 6\% and also predict higher radii for the binary components of GW170817 by 200-300m compared to the URs used previously in the literature. We further perform injection and recovery studies on simulated events with different EoSs in $\rm A+$ detector configuration as well as with third generation (3G) Einstein telescope. In agreement with the literature, we find that neglecting f-mode dynamical tides can significantly bias the inferred NS properties, especially for low mass NSs. However, we also find that the impact of the URs is within statistical errors.

\end{abstract}

                              
\maketitle


\section{Introduction}  
\label{sec:intro} 
The density in the core of a neutron star (NS) surpasses the nuclear saturation density ($n_0 \sim 2 \times 10^{14}\  \rm g\ cm^{-3}$)  and even the highest density that can be achieved in terrestrial experiments. It is argued that strangeness in the form of hyperons, meson condensates, or even deconfined quark matter may appear at such high densities, affecting several NS observable properties~\cite{VidanaEPJA}. The behavior of ultra-dense NS matter is still largely unknown, and therefore, NSs provide a natural laboratory to study nuclear matter under extreme conditions such as high magnetic field and rotation~\cite{Lattimer2004}.
\\

The  NS macroscopic observables  relate to the microscopic NS physics through the  pressure density relationship or the equation of state (EoS).  The two main approaches used to model nuclear EoS are (i)  microscopic or {\it ab-initio} ~\cite{OertelRMP,Sabatucci2022,Hebeler_2013,APR,STONE2007587} and (ii) phenomenological (effective theories with parameters fitted to reproduce saturation nuclear properties) ~\cite{MACHLEIDT19871,Haidenbauer}. Yet another approach  is to construct empirical fits rather than microphysics-based EoS. Where the high-density behavior of EoS is described by polytropic models (pressure is proportional to a power of density) either using piece-wise polytropes~\cite{Read2009} or using the spectral decomposition method~\cite{Spectral}. One can also infer the NS EoS from observations using a nonparametric description of NS EoS \cite{Essick2020} or can even use a hybrid approach to describe the NS EoS~\cite{Biswas2021}.
\\

On the observational side, NSs are observed at multiple wavelengths in current generation electromagnetic telescopes. The masses of these objects are typically well measured in binary systems, whereas the measurement of radius involves larger uncertainty. However, the recently launched NICER  (Neutron Star Interior Composition Explorer) mission ~\cite{NICER} has improved radius measurement and can further improve the estimation of the NS radius upto 5\%~\cite{Miller2019,Riley2019}. Additionally, binary neutron star (BNS) mergers are sources of gravitational waves (GW). In a BNS merger, tidal deformation of the NSs carries information about NS radii and is used to constrain the NS EoS in combination with mass measurements from the same binary~\cite{Agathos2015,Hinderer2010,AbbottPRL119,AbbottPRL121,AbbottPRX}. The ground-breaking BNS merger GW170817 ~\cite{AbbottAJL848,2019ApJ875160A,AbbottPRX} also observed electromagnetic (EM) counterparts in addition to GWs, and hence opened a new window in multi-messenger astronomy.
\\

The tidal deformation of the stars in a BNS contributes to  phase of the GW signal starting at the fifth Post Newtonian (PN) order. In general, 
tidal deformability (for electric type) of order `$l$' ($\Lambda_l$) appears at $2l+1$ Post Newtonian (PN) order. The major contribution comes from the  quadrupolar tides, followed by higher-order octupolar ($\Lambda_3$)
and hexadecapolar ($\Lambda_4$) terms~\cite{Hinderer2010,Henry2020}. Recent waveform models have been updated to consider the effect of the higher order tidal parameters  $\Lambda_3$, $\Lambda_4$ and also the magnetic type deformation (quadrupolar magnetic  $\Sigma_2$ in the GW phase ~\cite{Henry2020,Rossella2021}. In most analyses, the tides are considered adiabatic, \textit{ie.} the GW frequency is much lower than the NS resonant oscillation mode frequency. However, recent efforts largely support the excitation of  NS fundametal modes (f-modes) in the late inspiral ~\cite{Hinderer2016,Andersson2018,Ma2020,Andersson2021,Steinhoff2021,Gamba2022}. In a frequency domain model, the dynamical tidal contribution to the GW phase due to the excitation of f-modes appears at 8 PN order ~\cite{Schmidt2019,Hinderer2010} and depends upon the f-mode frequency. Since the additional tidal deformability parameters and the f-mode dynamical phase contribute at high PN orders, they are often dropped while constructing waveform approximants.  However, the f-mode dynamical tidal correction has been shown to significantly affect the inference of NS properties from a binary due to its resonance behaviour~\cite{Pratten2022, Williams2022}.
\\

Previous works have suggested that there exist EoS-insensitive universal relations (UR) between $\Lambda_2$ and each of $\Lambda_3$, $\Lambda_4$, and  $\Sigma_3$ ( hereafter referred as `multipole Love' relation) ~\cite{Yagi,YAGI20171,Yagi_2018}. Similarly, URs exist between the f-mode frequency and tidal parameters~\cite{Chan2014} ( hereafter referred as `f-Love' relation). This allows the additional contribution due to higher order tidal parameters or f-mode parameters to be considered without having to actually sample over them in a parameter estimation run, reducing the search parameter space. Additionally, there exists UR involving stellar compactness and quadrupole tidal parameters, which can be used to infer NS radius from the measured mass and quadrupole tidal parameters from a BNS event~\cite{Maselli}. In this work, we will only be interested in the URs mentioned above.   There also exist URs involving the combination of $\Lambda_2$ and mass ratio $q$ of the binaries, known as `binary Love relations,' as well as URs involving NS spin which reduces effort in measuring the NS tidal  parameter (and NS radius) in a BNS event~\cite{Yagi_2016,Chatziioannou2018,Kumar2019,Godzieba2021,YagiYunes2013,Gagnon2018,Carson2019}~\footnote{Recently a new EoS insensitive  approach involving the NS mass and tidal deformability is proposed to constrain NS  properties from a GW event~\cite{Biswas2022}. }. 
\\

The existing URs are employed by considering a few theoretical EoS models and some of the selective EoSs are now incompatible with the current astrophysical constraints~\cite{Yagi,Yagi_2018,Chan2014,Maselli}. Also, the existing URs do not consider the EoS uncertainties resulting from the nuclear parameters. In this work, we improve the multipole Love and f-Love URs relevant for GW astronomy and investigate the impact of the updated URs by analyzing the BNS event GW170817 and performing several injection and recovery studies with the future GW detector configurations. This work is organized in the following way. In \cref{sec:method} we discuss the choices of Eos, followed by the description of the methodology to solve for multipole tidal parameters and to solve for the NS f-mode characteristics. We compile our results in \Cref{sec:results} and summarise our conclusions in ~\Cref{sec:conclusion}~.
\\

\section{Method}\label{sec:method}
\subsection{Choices of Equations of State}
\label{subsec:eos}

The EoS is essentially the relation between the pressure ($p$) and density ($\rho$) (or energy density $\epsilon$), i.e., $p=p(\epsilon)$. As described in ~\Cref{sec:intro}, different physical motivations develop a diverse family of EoSs depending upon the physical descriptions of the  NS matter. In this work, we consider a wide range of EoSs based upon different physical descriptions. They are discussed below:

\paragraph{ Relativistic Mean Field (RMF) Models:} RMF models are phenomenological models where baryon-baryon interaction is mediated via exchange of mesons. The  Lagrangian density describes the interaction between baryons through the exchange of mesons. The complete description of the Lagrangian density for nucleonic ($npe\mu$) and nucleon-hyperon ($npe\mu Y$) mattered NS, and hence the EoS within the RMF model considered in this work can be found in ~\cite{Pradhan2021}. The parameters of the RMF model are calibrated to the nuclear and hypernuclear parameters at saturation: nuclear saturation density ($n_0$), the binding energy per nucleon  ($E/A$ or $E_{\rm sat}$), incompressibility  ($K$), the effective nucleon mass ($m^*$), symmetry energy ($J$) and slope of symmetry energy ($L$) at saturation. Hyperon coupling constants are fixed using hyperon nucleon potential depths ($U_Y$) or using symmetry properties. We consider the total uncertainty ranges in the saturation parameters resulting from nuclear and hypernuclear experiments, summarized in ~\Cref{tab:rmf_parameters}~\cite{Ghosh2022b}.
\\

\begin{table*}[ht]
    \centering
\begin{tabular}{c c c c c c c c c c}
\hline
    Model & $n_0$ & $E_{sat}$& $K$& $J$ & $L$& $m^*/m_N$ & $U_{\Sigma}$ & $U_{\Xi}$\\
      &($fm^{-3}$) & (MeV) & (MeV) & (MeV) & (MeV)&  & MeV & MeV\\
\hline
 \hline
    
    RMF~\cite{Pradhan2021}&[0.14, 0.17] & [-16.5, -15.5] & [200, 300] & [28, 34] & [40, 70] & [0.55, 0.75] & [0, +40] & [-40,0]\\
    
    \hline
\end{tabular}
\caption{Range of nuclear and hypernuclear saturation parameters considered in this work.
Meson and nucleon masses are fixed at $m_{\sigma}=550 \ \rm{ MeV}$, $m_{\omega}=783\  \rm{ MeV}$, $m_{\rho}=770\  \rm{MeV}$, $m_{\sigma^*}=975\  \rm{MeV}$, $m_{\phi}=1020\  \rm{MeV}$ and $m_N=939\  \rm{MeV} $. Masses of the hyperons are fixed from~\cite{PDG2020}. Note that we fix the $\Lambda$ hyperon potential depth ($U_{\Lambda}$=-30 MeV). }
\label{tab:rmf_parameters}
\end{table*}
 
\begin{figure*}[htbp]
    \begin{subfigure}{.45\textwidth}
  \centering
  \includegraphics[width=\linewidth]{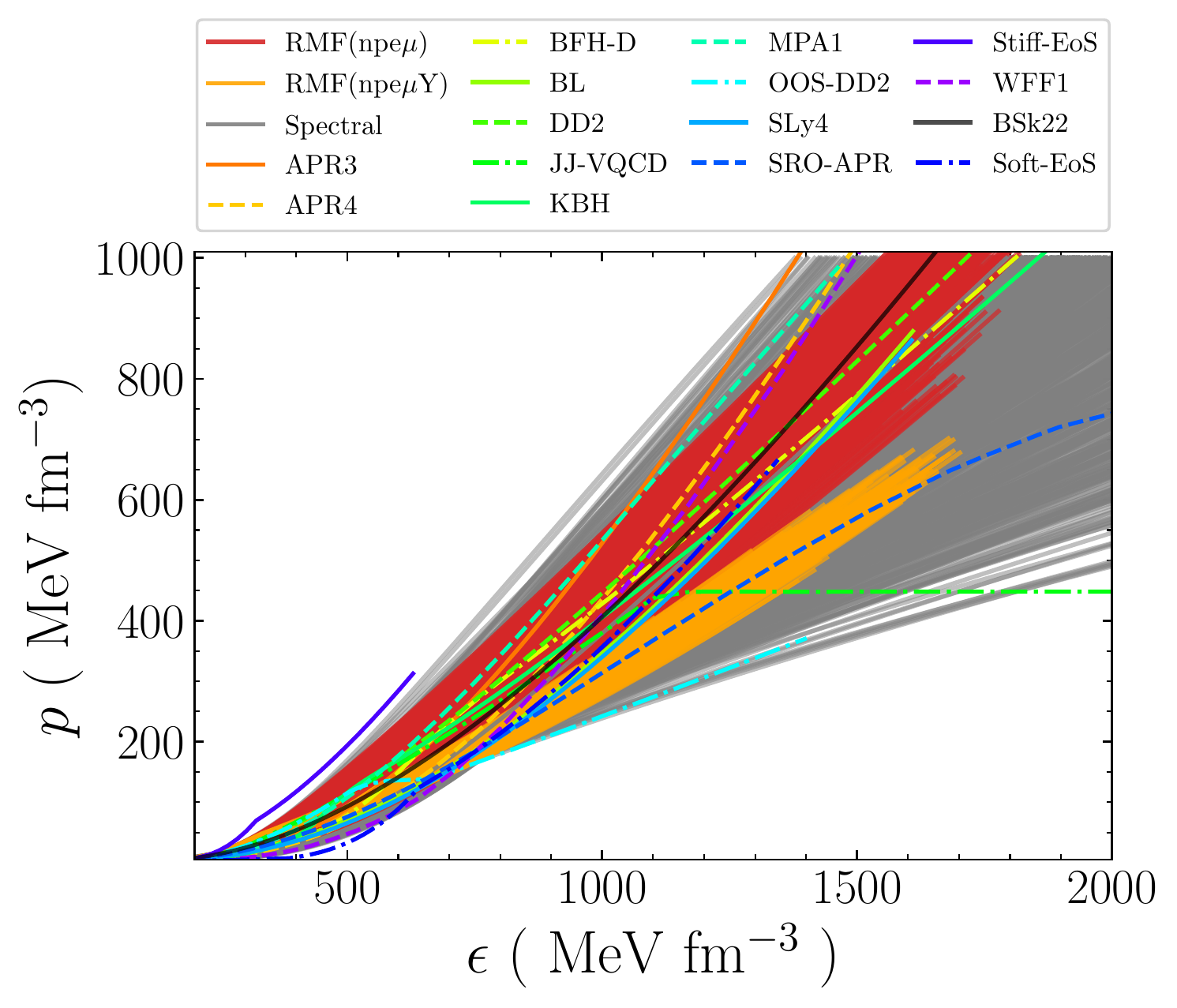}
  \caption{}
  \label{fig:eos}
    \end{subfigure}
    \begin{subfigure}{0.45\textwidth}
      \includegraphics[width=\linewidth]{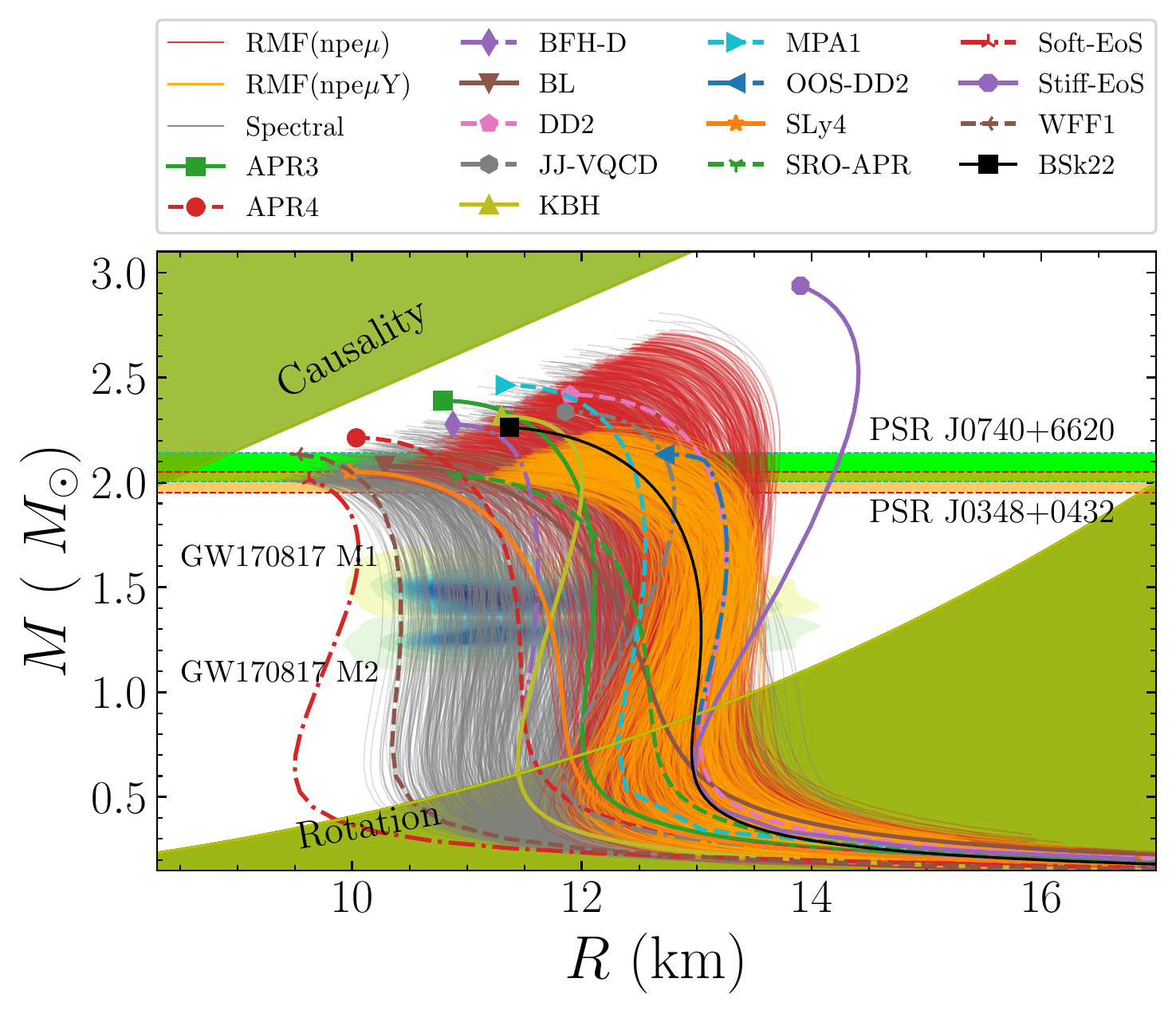}
  \caption{}
  \label{fig:mr}
    \end{subfigure}
    \caption{(\subref{fig:eos}) EoSs used in this work. (\subref{fig:mr}) Mass radius relation  corresponding to the EoSs used in this work. For each EoS, the corresponding M-R relationship is shown up to the maximum possible stable mass. Horizontal bands correspond to masses $M=2.072^{+0.067}_{-0.066} M_{\odot}$ of PSR J0740+6620 ~\cite{riley2021} and $M=2.01^{+0.04}_{-0.04} M_{\odot}$ of PSR J0348$+$0432 ~\cite{Antoniadis2013}. The mass radius estimates of the two companion neutron stars in the merger event GW170817~\cite{AbbottPRL121} are shown by the shaded area labeled with GW170817 M1 (M2) \footnote{\url{https://dcc.ligo.org/LIGO-P1800115/public}}. }
    \label{fig:eos_mr}
\end{figure*}

\paragraph{ Selective EoSs:} Along with the RMF  models with nucleonic and hyperonic mattered EoSs, we consider many realistic EoSs. The realistic EoSs are taken either from CompOSE database  ~\cite{Compose1,Compose2,Compose3} or from LALSimulation ~\cite{lalsuite}. The considered realistic EoSs are APR4, APR3~\cite{APR,Douchin2001}, SLy4~\cite{Gulminelli2015}, BL~\cite{BL2018}, DD2(GPPVA)~\cite{Grill2014}, SRO-APR~\cite{SROAPR}, BSk22 ~\cite{Bsk22_2018,Bsk22_2022}, WFF1 ~\cite{WFF1}, MPA1 ~\cite{MPA1}. We also consider the Soft and Stiff EoS from ~\cite{Hebeler_2013}. In  ~\Cref{fig:eos}, though the Stiff-EoS terminates at earlier energy density, it is sufficient to reach the maximum stable NS mass the EoS model can reproduce ~\cite{Hebeler_2013}. To capture the hypothesis of a deconfined quark phase in the interior of the NS core, we account for some realistic hybrid EoSs containing the quark phase in the interior. The selective hybrid EoS models are BFH-D ~\cite{Baym_2019,TOGASHI2017}, KBH ($\rm QHC21-AT$)~\cite{KBH_2022}, JJ-VQCD ~\cite{JJVQCD} and OOS-DD2(FRG) ~\cite{Otto2020}. We do not consider the EoSs regarding quark stars only, as they might deviate from the universal behavior and leave them for a separate investigation~\cite{Yagi,Wen}.
\\

\paragraph{ Spectral Decomposition: } To span the EoSs with empirical fit formalism, we consider  the four parameters spectral decomposition method
developed in ~\cite{Spectral}. In spectral decomposition, the adiabatic index ($\Gamma$) of the EoS is spectrally decomposed onto a set of polynomial basis functions and expressed as,
\begin{equation}
    \Gamma (p)=\exp{\l(\sum_k \gamma_k \l[ln{(p/p_0)}\r]^k\r)}
\end{equation}

where $\gamma_k$ is the expansion coefficient and $p_0$ is the reference pressure where the high-density EoS is stitched to the low-density crustal EoS. The EoS can then be generated by integrating the relation,
\begin{equation}
    \frac{d\epsilon}{dp}=\frac{\epsilon+p}{p\Gamma (p)}
\end{equation}
which can be reduced to,
\begin{equation}
    \epsilon(p)=\epsilon(p_0)+ \frac{1}{\mu(p)}\int_{p_0}^{p}\frac{\mu(p^{\prime})}{\Gamma(p^{\prime})} dp^{\prime}
\end{equation}
where, $$\mu(p)=\exp \l( -\int_{p_0}^{p}\frac{dp^{\prime}}{p^{\prime} \Gamma(p^{\prime})} \r)$$~.

We fix the low-density EoS to SLy EoS ~\cite{SLy} and stitch the high-density EoS at a density below half of the saturation density such that the NS macroscopic properties will not be affected significantly. We generate spectral decomposed  EoSs as implemented in  
LALSimulation~\cite{lalsuite} and consider the ranges for spectral indices ($\gamma_k$) from ~\cite{AbbottPRL121,Carney2018}.

Before proceeding with any further calculations, we ensure that each EoS satisfies the required physical conditions, such as thermodynamic stability ($dp/d\epsilon>0$), causality ($\sqrt{dp/d\epsilon}\leq 1$) and the monotonic behaviour of pressure ($dp/d\rho>0$ and $d\epsilon/d\rho>0$). We additionally impose the constraint that the EoS must be able to produce a $2M_{\odot}$ stable NS and the tidal deformability of a $1.4M_{\odot}$ is less than 800 (i.e, ${\Lambda}_{1.4M_{\odot}}\leq 800$)~\cite{AbbottPRX,AbbottPRL121}. EoSs and the corresponding mass-radius relations used in this work are displayed in ~\Cref{fig:eos} and ~\Cref{fig:mr} respectively.

\subsection{Macroscopic Structure and Tidal Deformabilities}
\label{subsec:nsproperties}
For a given EoS, NS mass ($M$) and radius ($R$) are obtained by solving the  Tolman-Oppenheimer-Volkoff (TOV) equations. The vanishing of pressure at the surface of NS ($p(R)=0$) provides the stellar radius $R$ and the stellar mass $M=m(R)$. TOV equations corresponding to a static and spherically symmetric metric~\eqref{eqn:metric} are summarised in Eq.~\eqref{eqn:tov}~\cite{Tolman,Oppenheimer}.
\begin{equation}
     ds^2=-e^{2\Phi (r)}dt^2+e^{2\lambda (r)}dr^2+r^2 d\theta^2+r^2\sin^2{\theta} d\phi^2 \label{eqn:metric}
 \end{equation}
\begin{eqnarray}\label{eqn:tov}
     \frac{dm(r)}{dr}&=& 4\pi r^2 \epsilon (r)  \label{tov1} \ , \nonumber\\
    \frac{dp(r)}{dr}&=&- \left[p (r)+\epsilon (r)\right]\frac{m (r)+4\pi r^3p (r)}{r (r-2m (r))} \label{tov2} \nonumber\ ,\\
    \frac{d \Phi(r)}{dr}&=&\frac{-1}{\epsilon(r)+p(r)}\frac{dp}{dr} \label{phi} \nonumber\ ,\\
     e^{2\lambda (r)}&=& \frac{r}{r-2m(r)} \,.\label{lambda}~.
\end{eqnarray}
The NS can be tidally deformed in a binary system due to mutual gravitational interaction. The tidal field can be decomposed to an electric ($\mathcal{E}_{ij}$) and magnetic ($\mathcal{M}_{ij}$) component, leading to the induction of mass multipole moment ($\mathcal{Q}_{ij}$) and current multipole moment ($S_{ij}$). The gravitoelectric tidal deformability ($\lambda_l$) and gravitomagnetic tidal deformability ($\sigma_l$) of order $l$ can be defined as~\cite{Hinderer_2008},
\begin{eqnarray}
    \mathcal{Q}_{ij}&=&\lambda_l \mathcal{E}_{ij}\ , \nonumber \\
    S_{ij}&=&\sigma_l \mathcal{M}_{ij}
\end{eqnarray}

\begin{eqnarray}
     k_l&=&\frac{(2l-1)!!}{2}\  \lambda_l\ ,\nonumber\\
     j_l&=&{4(2l-1)!!}\ \sigma_l~.
     \label{eq:Lovenumber}
\end{eqnarray}

The electric Love number ($k_l$) and magnetic Love number ($j_l$) relate to the tidal deformability parameters as given in Eq.~\eqref{eq:Lovenumber}. The valuable parameters that can be determined from a GW signal are dimensionless tidal deformability. The dimensionless gravitoelectric  tidal deformability parameter ($\Lambda_l$) and   gravitomagnetic  tidal deformability parameter ($\Sigma_l$) can be expressed using the corresponding tidal Love numbers and stellar compactness ($C=M/R$) as~\cite{Perot_2021}:

\begin{eqnarray}
     \Lambda_l&=&\frac{2}{(2l-1)!!}\frac{k_l}{C^{2l+1}},\ \nonumber \\
     \Sigma_l&=&\frac{1}{4(2l-1)!!}\frac{j_l}{C^{2l+1}}~.
\end{eqnarray}
For computing the electric and magnetic  Love numbers, one needs to integrate additional set of differential equations~\cite{Perot_2021} along with the TOV equations. We follow the methodology developed in ~\cite{Perot_2021} to solve for the Love numbers. We test our numerical scheme by reproducing the Love numbers corresponding to selective EoSs used in ~\cite{Perot_2021}. We provide the electric tidal deformability up to order `$\ell=4$' and magnetic tidal deformability up to order `$\ell=3$'.

\subsection{Finding f-mode oscillation characteristics}
\label{subsec:fmode}
A neutron star can have several quasi-normal modes depending upon the restoring force. Mode characteristics (frequency and damping time) contain information about the NS interior. Hence, the determination of mode parameters can be employed to constrain the NS interior or EoS. In our previous work ~\cite{Pradhan2022}, we had shown that depending upon the EoS, the Cowling approximation can overestimate the quadrupolar f-mode frequency up to about 30\%. As recent efforts are going on to improve the gravitational waveform models with consideration of excitation of f-modes which depends on the mode frequency~\cite{Schmidt2019}, one should consider the general relativistic formalism to find the mode characteristics. We use the direct integration method developed in ~\cite{Thorne1967,Detweiler85,Sotani2001,Pradhan2022} to find the NS f-mode frequency. In short, the coupled equations for perturbed metric and fluid variables are integrated within the NS interior with appropriate boundary conditions~\cite{Detweiler85}. Outside of the NS, fluid variables are set to zero, and then the Zerilli's wave equation ~\cite{Zerilli} is integrated to far away from the star. Then a search is carried out for the complex f-mode frequency ($\omega=2\pi f+\frac{i}{\tau_f}$) for which one has only outgoing wave solution to the  Zerilli's equation at infinity. The real part of $\omega$ represents the f-mode angular frequency, and the imaginary part represents the damping time. For finding the mode characteristics, we use the numerical methods developed in our previous work~\cite{Pradhan2022}.

\section{Results}
\label{sec:results}
\subsection{Multipole Love and f-Love Relations}
The universality among higher-order electric  tidal deformability parameters ($\Lambda_{\ell \geq 3}$ ) or the magnetic deformability parameter ($\Sigma_{\ell \geq 2}$) with the quadrupolar tidal deformability parameter ($\Lambda_{\ell =2}$) were first discussed in ~\cite{Yagi,Yagi_2016} with few selective EoSs. The universality of compactness ($C$) with $\Lambda_2$ initially introduced in ~\cite{Maselli} and the UR involving f-mode frequency and tidal deformability parameters are  introduced  in ~\cite{Chan2014}.   Original URs from ~\cite{Yagi,Yagi_2018,Chan2014,Maselli} are expressed  using a polynomial fit of the following form (the order of the polynomial differs in different works). 

\begin{eqnarray}
    P=\sum_{k=0}^{6} a_k \l[\ln{(\Lambda_2)}\r]^k, \label{eqn:tidal_fit}
\end{eqnarray}
where, 
\begin{eqnarray*}
    P&=&\{ \ln(\Lambda_3),\ \ln(\Lambda_4),\ \ln(|\Sigma_2|),\ \ln(|\Sigma_3|),\ C, \nonumber \\
    & & \ M\omega_2,\ M\omega_3,\  M\omega_4\}
\end{eqnarray*}

In recent works~\cite{Godzieba2021,Godzieba2021b}, the universal relations for electric type deformability were updated by considering the phenomenological piece-wise polytropic EoSs. In different waveform models, the correction on the tidal phase includes the electric tidal parameter up to order $\ell \leq4$ and for magnetic deformation up to order $\ell \leq3$. Hence, we provide the universal relations for $\Lambda_3$ and $\Lambda_4$ with $\Lambda_2$ for electric type, whereas for magnetic type, we provide the universal relation of  $\Sigma_2$ and $\Sigma_3$  with $\Lambda_2$. 
From the tidal deformability and mass obtained from a binary system, one can infer the NS radius ($R$) by using the EoS-independent relation that exists between stellar compactness ($C=M/R$), and tidal deformability parameter ~\cite{Maselli,Godzieba2021,shashank2021,Biswas2022,Das2022,Kashyap2022}\footnote{The systematic errors that could be occurred in the inferred NS properties due to the choice of binary Love relation or $C-\Lambda_2$ UR are recently discussed in ~\cite{Godzieba2021,Kashyap2022}.}. There are  EoS-independent relations, which involve the tidal deformability of both binary NSs and binary parameters ( like $q=m_1/m_2$). The universal relation involving the binary parameters and the $C-\Lambda_2$ relation is also used to infer the NS radii~\cite{Yagi_2016,Yagi_2016b,Chatziioannou2018,Godzieba2021}. Though binary relations reduce the parameter space of BNS search parameters,  the inferred NS parameters are model dependent~\cite{Kastaun2019}. 

\begin{figure*}[htbp]
    \centering
    \begin{subfigure}{.45\textwidth}
  \centering
  \includegraphics[width=\linewidth]{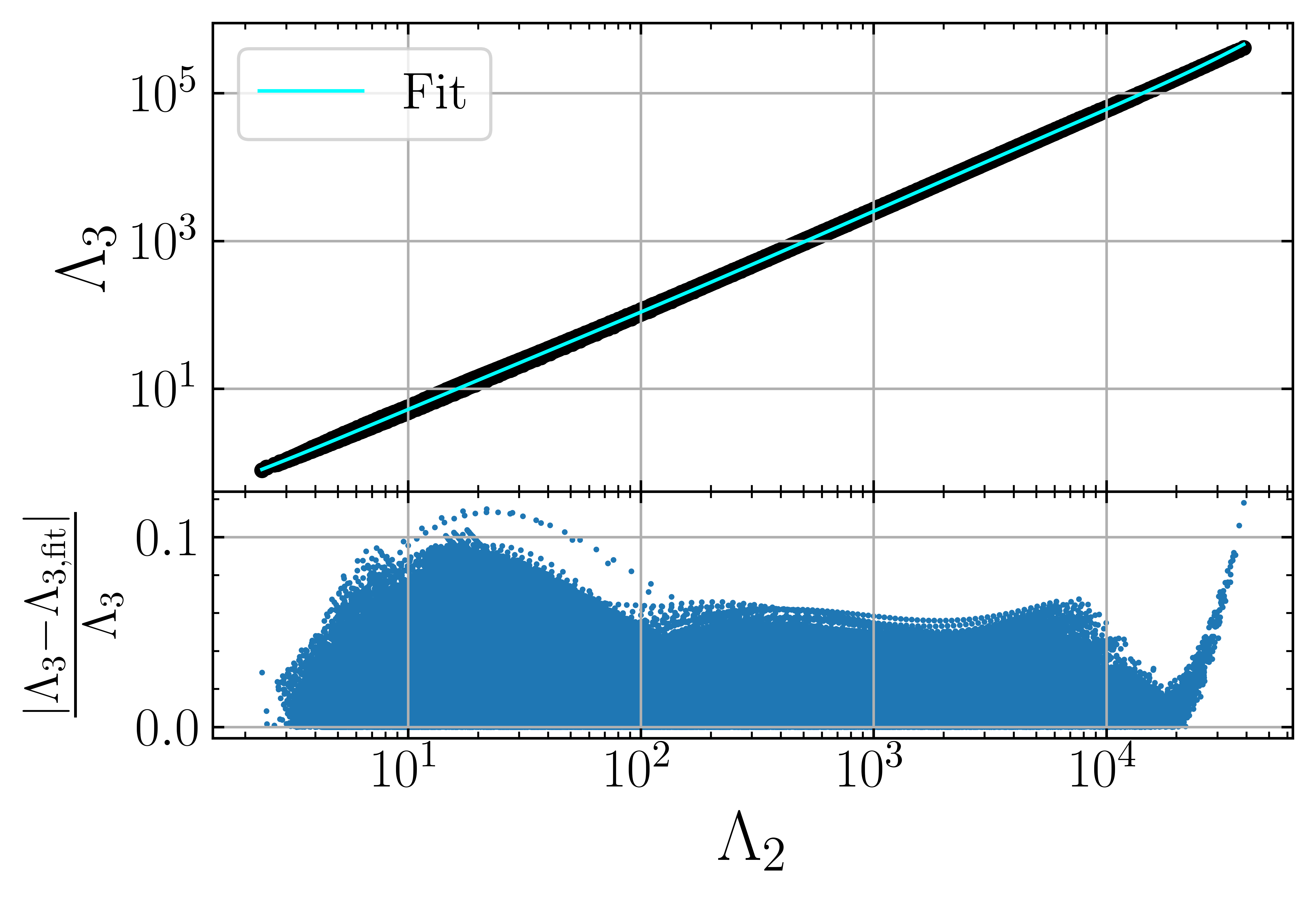}
  \caption{}
  \label{fig:tidal3_tidal2}
    \end{subfigure}
    \begin{subfigure}{0.45\textwidth}
      \includegraphics[width=\linewidth]{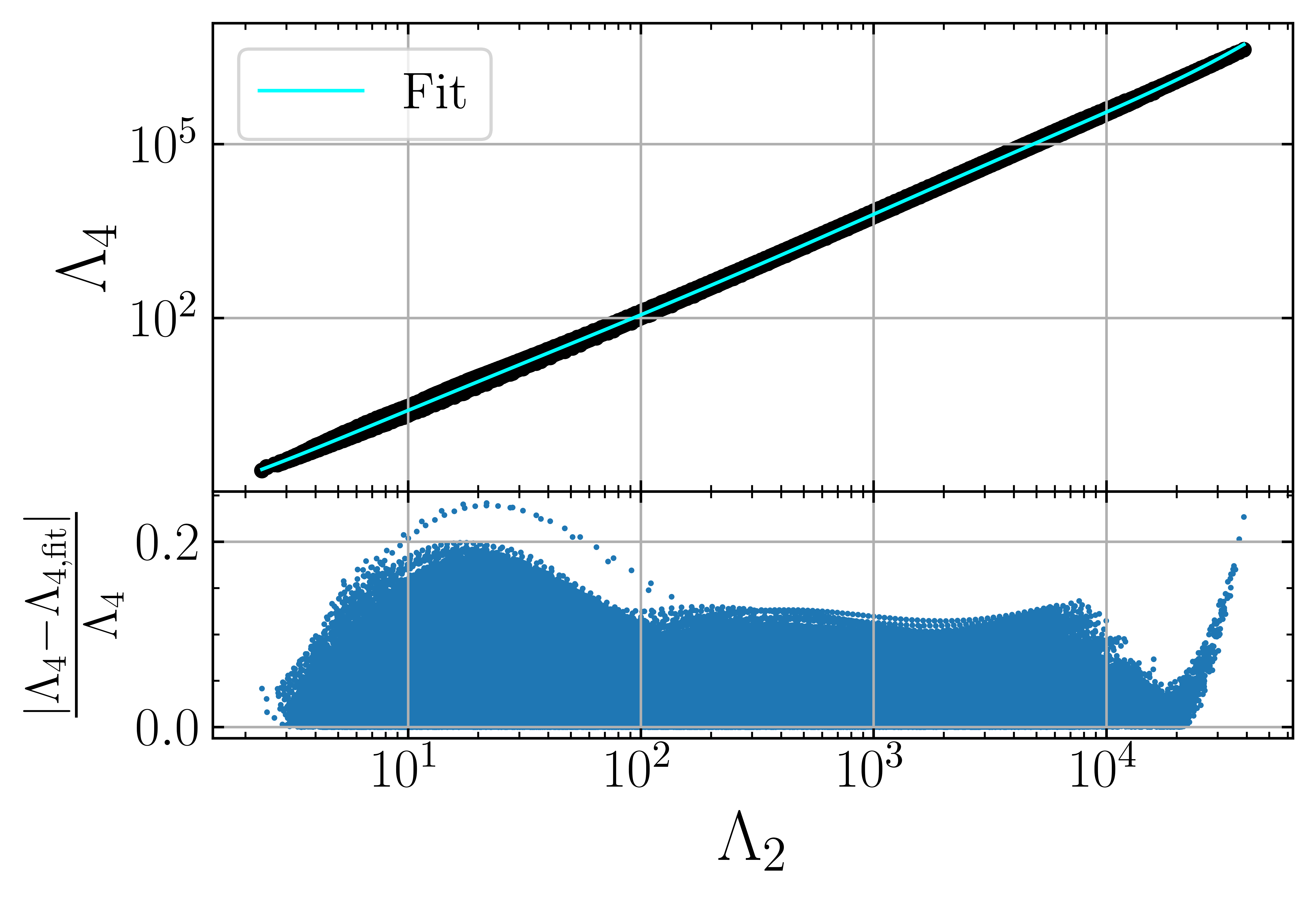}
  \caption{}
  \label{fig:tidal4_tidal2}
    \end{subfigure}
    \caption{Showing Universal relations of higher order tidal deformabilities with $\Lambda_2$ along with the relative errors corresponding to fit relation (~\subref{fig:tidal3_tidal2}) for $\Lambda_3$  and (~\subref{fig:tidal4_tidal2}) for  $\Lambda_4$.}
    \label{fig:multi_Love}
\end{figure*}

\begin{figure*}[htbp]
    \centering
    \begin{subfigure}{.45\textwidth}
  \centering
  \includegraphics[width=\linewidth]{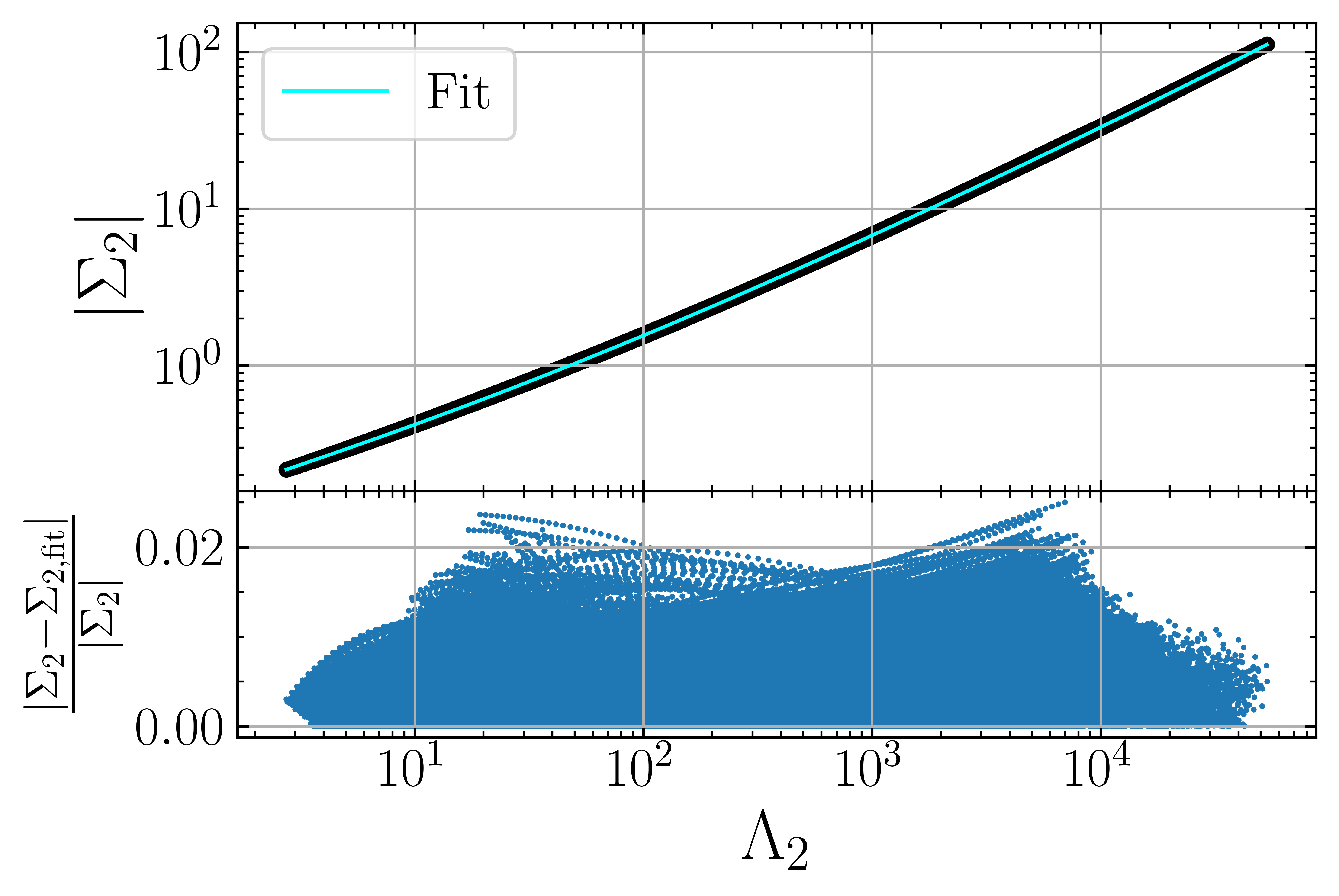}
  \caption{}
  \label{fig:sigma2_tidal2}
    \end{subfigure}
    \begin{subfigure}{0.45\textwidth}
      \includegraphics[width=\linewidth]{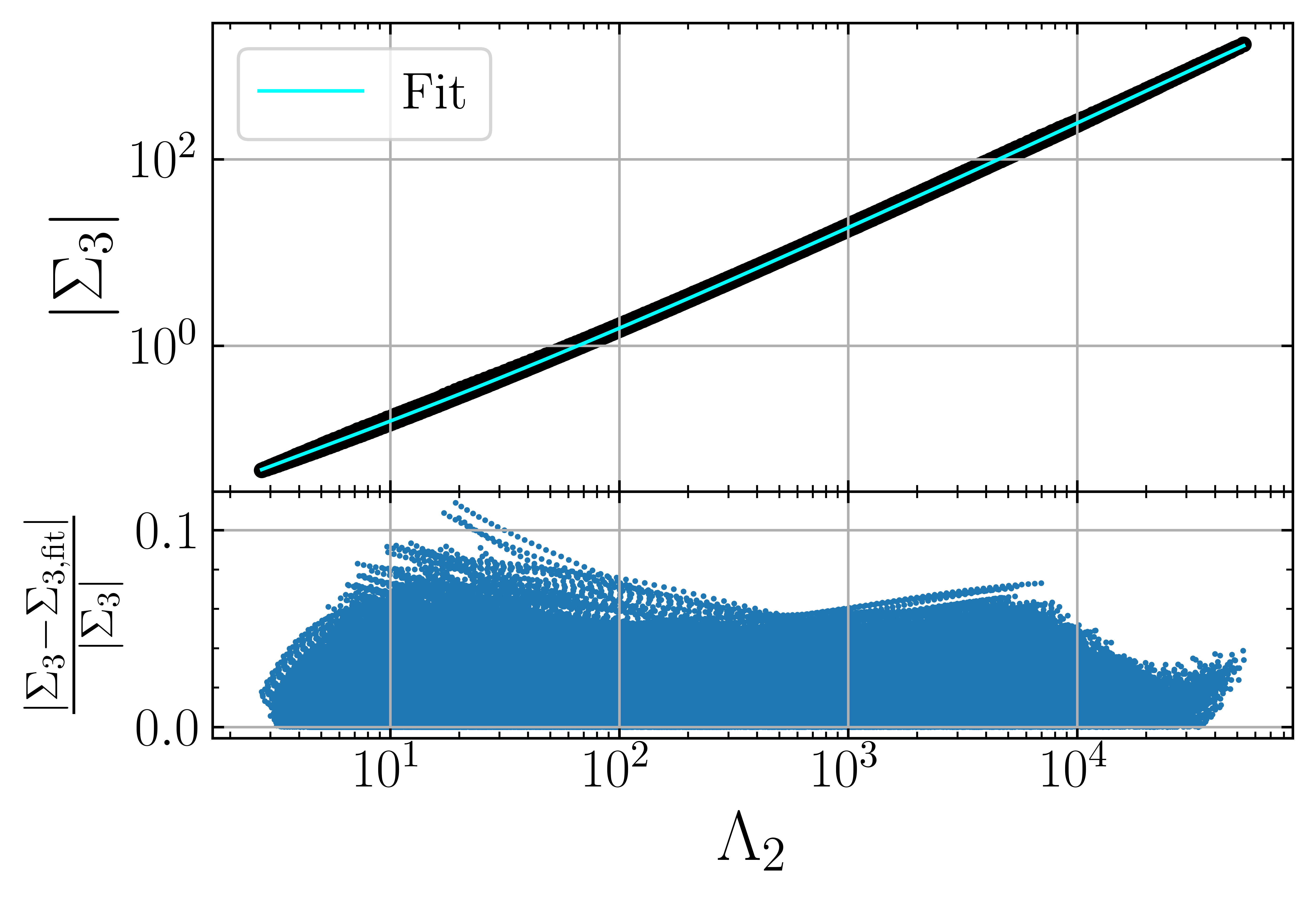}
  \caption{}
  \label{fig:sigma3_tidal2}
    \end{subfigure}
    \caption{Showing Universal relations of  tidal deformabilities of magnetic type with $\Lambda_2$ along with the relative errors corresponding to fit relation (~\subref{fig:sigma2_tidal2}) for $\Sigma_2$  and (~\subref{fig:sigma3_tidal2}) for  $\Sigma_3$. }
    \label{fig:multi_Love_magnetic}
\end{figure*}

\begin{figure}[htbp]
    \centering
    \includegraphics[width=\linewidth]{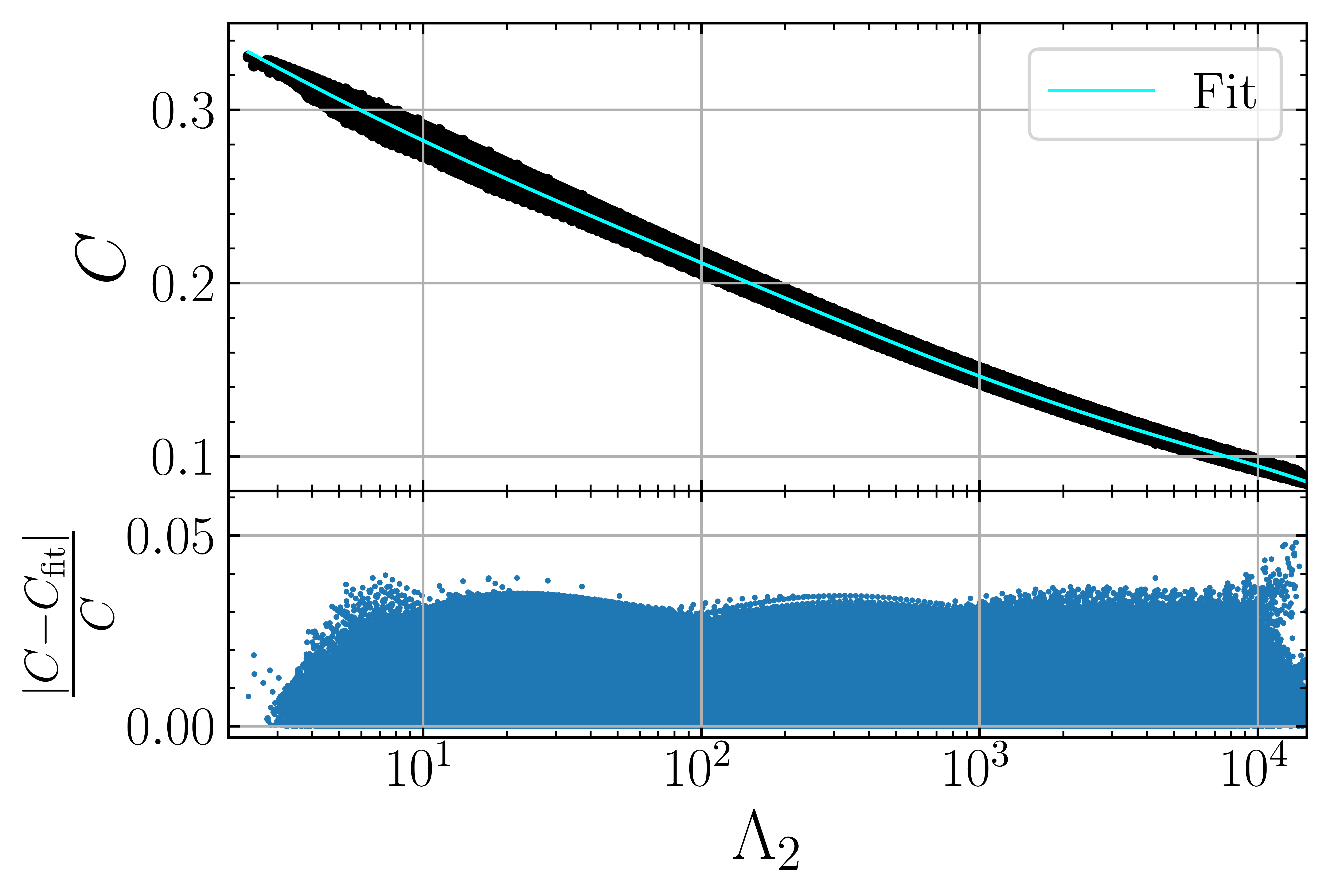}
    \caption{ ( Upper panel ) Universal relations among stellar compactness ($C$) and dimensionless quadrupolar tidal deformability ($\Lambda_2$). (Lower panel) Relative error corresponding to fit relation.}
    \label{fig:C_Love_rlation}
\end{figure}

As discussed in ~\Cref{subsec:eos}, we consider tentatively  6000 EoSs originating from different physical motivations. We fix the lower mass for an EoS by imposing the constraint resulting from the maximum rotation pulsar PSR J1748-2446~\cite{Jason2006,Lattimer2019} (also notice the rotation limiting curve in ~\Cref{fig:mr}). The maximum mass for an EoS is fixed at the maximum stable NS mass the EoS can produce. We then use the method described in ~\Cref{subsec:nsproperties} to obtain the multipole tidal parameters ($\Lambda_{\ell}$ and $\Sigma_{\ell}$). We obtain the multipole Love relations by solving the NS properties (with tidal parameters) for $\sim 1.2\times 10^6$ neutron stars. We produce the universal relation for multipole Love relations as a polynomial fit ~\eqref{eqn:tidal_fit} as introduced in ~\cite{Yagi}. Note that the original fit from Yagi ~\cite{Yagi} was a $4^{th}$ order polynomial which was then updated to a $6^{th}$ order polynomial for a larger data set in ~\cite{Godzieba2021}. In our data set, we notice that by updating the polynomial from a quartic polynomial to a $5^{th}$ order polynomial, the resulting goodness of fit improved significantly and then did not improve significantly after increasing the order of the polynomial. However, we provide the universal relations with a  $6^{th}$ order polynomial~\eqref{eqn:tidal_fit} to be consistent with the updated relations. Universal relations $\Lambda_3-\Lambda_2$, $\Lambda_4-\Lambda_2$, $\Sigma_2-\Lambda_2$ and $\Sigma_3-\Lambda_2$ are displayed in \Cref{fig:tidal3_tidal2,fig:tidal4_tidal2,fig:sigma2_tidal2,fig:sigma3_tidal2} respectively. The fit parameters for the multipole Love relations ~\eqref{eqn:tidal_fit}  involving tidal parameters obtained from this work and other works are tabulated in ~\Cref{tab:multipole_Love_fitparameters}. Similarly, for $C-\Lambda_2$ relation we provide the fit parameters in ~\Cref{tab:C_Love_fitparameters} and displayed the relation in ~\Cref{fig:C_Love_rlation}. Our multipole universal relations are valid for $ 2.3 \leq \Lambda_2\leq 4\times 10^{4} $ which is the essential  range for GW astronomy ( mostly in the range $\leq 10^4$) and for $C-\Lambda_2$, the relation is valid for $0.08\leq C \leq0.33$. 

Although the detection of f-modes from BNS would only be possible with third generation detectors~\cite{Williams2022}, the impact on the inferred NS parameters can be seen in the A+ detector configurations~\cite{Pratten2022,Gamba2022}. In the frequency domain, the phase correction due to the dynamical excitation of f-modes depends solely on the mode frequency~\cite{Hinderer2010,Hinderer2016,Schmidt2019,Pratten2020}. The universality of mass scaled f-mode angular frequency (\textit{ie.}, $M\omega$) with the tidal parameter ($\Lambda$) was first studied in the work of Chan \textit{et al.} ~\cite{Chan2014}, where the universality was explained by using the universal behavior of f-mode frequency with the moment of inertia ($I$), and the $I-Love-Q$ relations ~\cite{Chan2014,YagiYunes2013}. To avoid the numerical instabilities, we fix the lower mass $\sim 1M_{\odot}$ for each EoS despite the rotation limit while solving for the f-mode frequency. The proposed quartic polynomial fit of ~\cite{Chan2014} is updated to a 5th order fit in ~\cite{Sotani2021,Pradhan2022}. We provide the polynomial fit~\eqref{eqn:tidal_fit} up to 6th order  to be consistent with other multipole universal relations.  Our f-Love relations are valid for $2.3\leq \Lambda_2 \leq 4\times 10^4$.


\begin{table*}[ht]
    \centering\small\setlength\tabcolsep{.2em}
    
    \begin{tabular}{c c c c c c c c c c}
    \hline \hline
      Relation&Work&$a_0$   & $a_1$ & $a_2$ &$a_3$ &$a_4$ &$a_5$ &$a_6$ & Error [\%]\\
      & & & & & & & & & Max (90\%)\\
    \hline 
      & Yagi ~\cite{Yagi}&-1.15&1.18&2.51$\times 10^{-2}$&-1.31$\times 10^{-3}$ & 2.52 $\times 10 ^{-5}$&--&--&16 (10) \\
     $\Lambda_3 - \Lambda_2$ &Godzieba et al.~\cite{Godzieba2021}&-1.052& 1.165& 6.369$\times 10^{-3}$& 5.058$\times 10^{-3}$& -7.268$\times 10^{-4}$&3.749$\times 10^{-5}$& -6.803$\times 10^{-8}$&23 (16)\\
    &This Work&-1.163&  9.442$\times 10^{-1}$&  2.492$\times 10^{-1}$& -8.170$\times 10^{-2}$&1.374$\times 10^{-2}$& -1.117$\times 10^{-3}$&  3.494$\times 10^{-5}$&11 (7)\\
    \hline
     & Yagi~\cite{Yagi}&-2.45&1.43&3.951$\times 10^{-2}$&-1.81$\times 10^{-3}$ & 2.80$\times 10^{-5}$& -- & --&35 (20)\\
     $\Lambda_4 - \Lambda_2$& Godzieba et al. ~\cite{Godzieba2021} &-2.262& 1.383& 1.662$\times 10^{-3}$& 1.225$\times 10^{-2}$&-1.752$\times 10^{-3}$& 9.667$\times 10^{-5}$& -1.886$\times 10^{-6}$ & 44 (30)\\
    &This Work &-2.533&  1.050&  4.216$\times 10^{-1}$& -1.433$\times 10^{-1}$&2.461$\times 10^{-2}$& -2.027$\times 10^{-3}$&  6.396$\times 10^{-5}$& 24 (13)\\
    \hline 
    $|\Sigma_2| - \Lambda_2$& Yagi ~\cite{Yagi_2018} & -2.01 & +0.462 & 0.168&-1.58 $\times10^{-4}$& -6.03 $\times10^{-5}$ &- &- &32 (28)\\
    
     & This Work & -2.019&4.821$\times 10^{-1}$&7.609$\times 10^{-4}$&4.096$\times 10^{-3}$&-5.03$\times 10^{-4}$&2.643$\times 10^{-5}$&-5.192$\times 10^{-7}$&3 (1.5)\\
     \hline
     $|\Sigma_3 |- \Lambda_2$& This Work &-4.029&1.017&-8.218$\times 10^{-2}$&2.977$\times 10^{-2}$&-4.258$\times 10^{-3}$& 2.903$\times 10^{-4}$&-7.737$\times 10^{-6}$& 12 (5)\\
     \hline \hline
    \end{tabular}
    \caption{Values of the fit parameters $a_j$  found in this work  for the given equation~\eqref{eqn:tidal_fit}. The error column shows the maximum error (and the upper bound of  90\% errors) that a UR holds in the range $\Lambda_2\leq10^4$.}
    \label{tab:multipole_Love_fitparameters}
\end{table*}
\begin{table*}[ht]
    \centering\small\setlength\tabcolsep{.12em}
    
    \begin{tabular}{c c c c c c c c c c}
    \hline \hline
      Relation&Work&$a_0$   & $a_1$ & $a_2$ &$a_3$ &$a_4$ &$a_5$ &$a_6$ & Error [\%]\\
      & & & & & & & & & Max (90\%)\\
    \hline 
       & Maselli et al. ~\cite{Maselli} &0.371 &-3.91$\times 10^{-2}$& 1.056$\times 10^{-3}$& & & & &11 (8)\\
     $C - \Lambda_2$&Godzieba et al.~\cite{Godzieba2021}&0.3389&-2.293$\times 10^{-2}$&-5.172$\times 10^{-4}$&-2.449$\times 10^{-4}$& 5.161$\times 10^{-5}$& -3.03$\times 10^{-6}$ &5.841$\times 10^{-8}$&8 (5)\\
      &This Work &3.663$\times 10^{-1}$&-3.727$\times 10^{-2}$&-2.243$\times 10^{-3}$&1.941$\times 10^{-3}$& -4.491$\times 10^{-4}$&4.463$\times 10^{-5}$&-1.606$\times 10^{-6}$& 4 (2.5)\\
     \hline \hline
    \end{tabular}
    \caption{Values of the fit parameters for $C -\Lambda_2$ universal relation ~\eqref{eqn:tidal_fit}.  {The error column shows the maximum error (and the upper bound of  90\% errors) that a UR holds in the range $\Lambda_2\leq10^4$.}}
    \label{tab:C_Love_fitparameters}
\end{table*}

\begin{table*}[ht]
    \centering
    \begin{tabular}{c | c   c  c c c c c c}
    \hline \hline
      Work&Relation&$a_0$   & $a_1$ & $a_2$ &$a_3$ &$a_4$ &$a_5$ &$a_6$ \\
    \hline 
      &$M \omega_2 - \Lambda_2$&1.820$\times 10^{-1}$&-6.836$\times 10^{-3}$& -4.196$\times 10^{-3}$&5.215$\times 10^{-4}$&-1.857$\times 10^{-5}$&--&--\\
     Chan et al.~\cite{Chan2014} &$M \omega_3 -\Lambda_3$&2.245$\times 10^{-1}$&-1.500$\times 10^{-2}$&-1.412$\times 10^{-3}$& +1.832$\times 10^{-4}$& -5.561$\times 10^{-6}$&--&--\\
      &$M \omega_4 - \Lambda_4$&2.501$\times 10^{-1}$&-1.646$\times 10^{-2}$&-5.897$\times 10^{-4}$ & 8.695$\times 10^{-5}$& -2.368$\times 10^{-6}$ & -- &--\\
      \hline
      &$M \omega_2 - \Lambda_2$&1.820$\times 10^{-1}$& -6.665$\times 10^{-3}$& -4.212$\times 10^{-3}$&  4.724$\times 10^{-4}$&
        -1.030$\times 10^{-6}$& -2.139$\times 10^{-6}$&  8.763$\times 10^{-8}$\\
      This Work &$M \omega_3 -\Lambda_2$& 2.389$\times 10^{-1}$& -7.869$\times 10^{-3}$& -7.213$\times 10^{-3}$&1.539$\times 10^{-3}$&-1.908$\times 10^{-4}$&1.414$\times 10^{-5}$&-4.434$\times 10^{-7}$\\
      &$M \omega_4 - \Lambda_2$&2.863$\times 10^{-1}$&-1.044$\times 10^{-2}$&-8.90$\times 10^{-3}$& 2.189$\times 10^{-3}$&        -3.112$\times 10^{-4}$&2.478$\times 10^{-5}$&-7.975$\times 10^{-7}$\\
     \hline \hline
    \end{tabular}
    \caption{Values of the fit parameters for mass scaled angular frequency ($M\omega_{\ell}$) and $\Lambda_2$ found in this work  for the given equation~\eqref{eqn:tidal_fit}~.}
    \label{tab:f_Love_fitparameters}
\end{table*}
\begin{figure}[htbp]

\begin{subfigure}{\linewidth}
\includegraphics[clip,width=\linewidth]{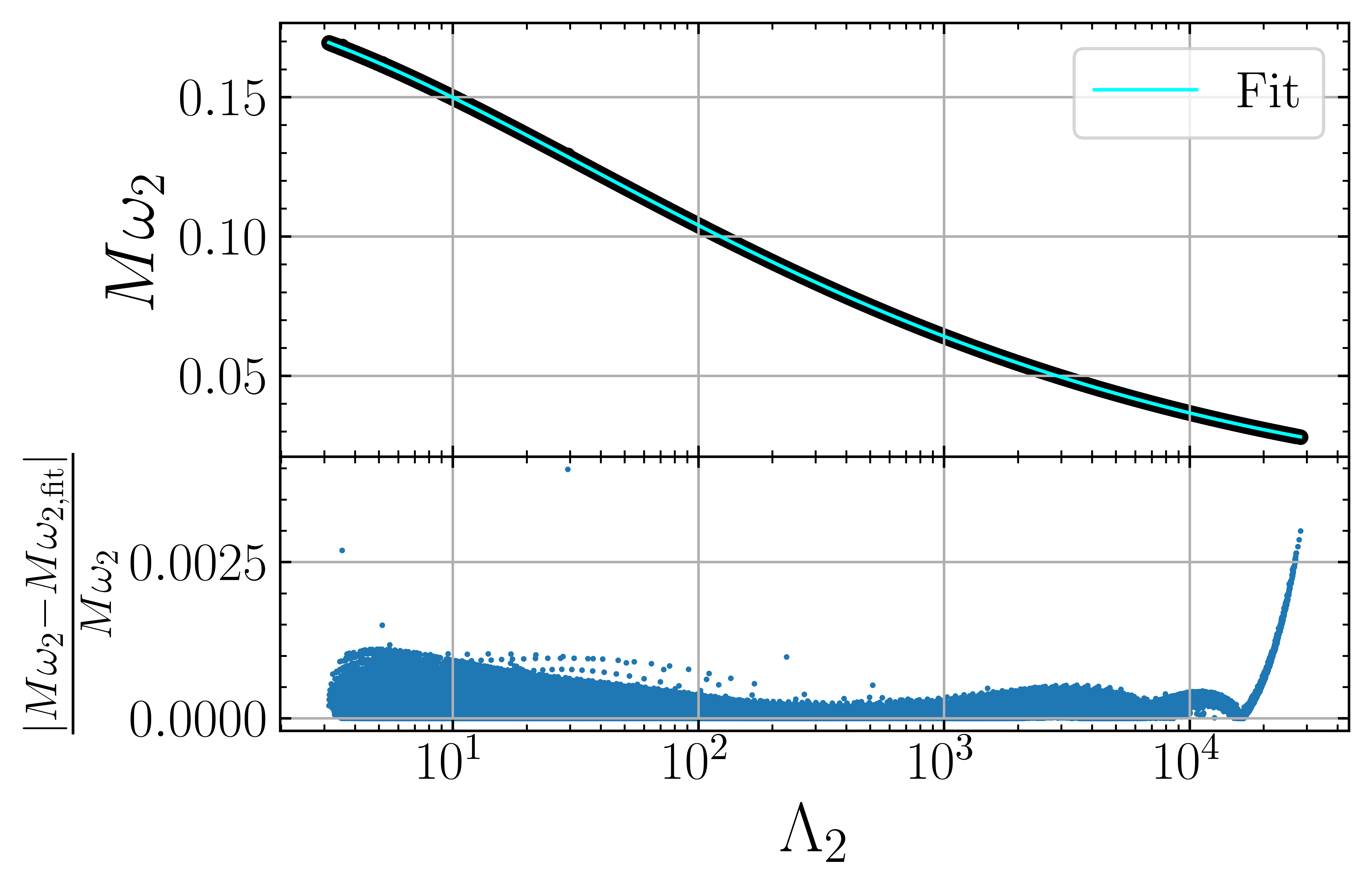}
\caption{}
\label{fig:f_Love_leq2}
\end{subfigure}

\begin{subfigure}{\linewidth}
\includegraphics[clip,width=\linewidth]{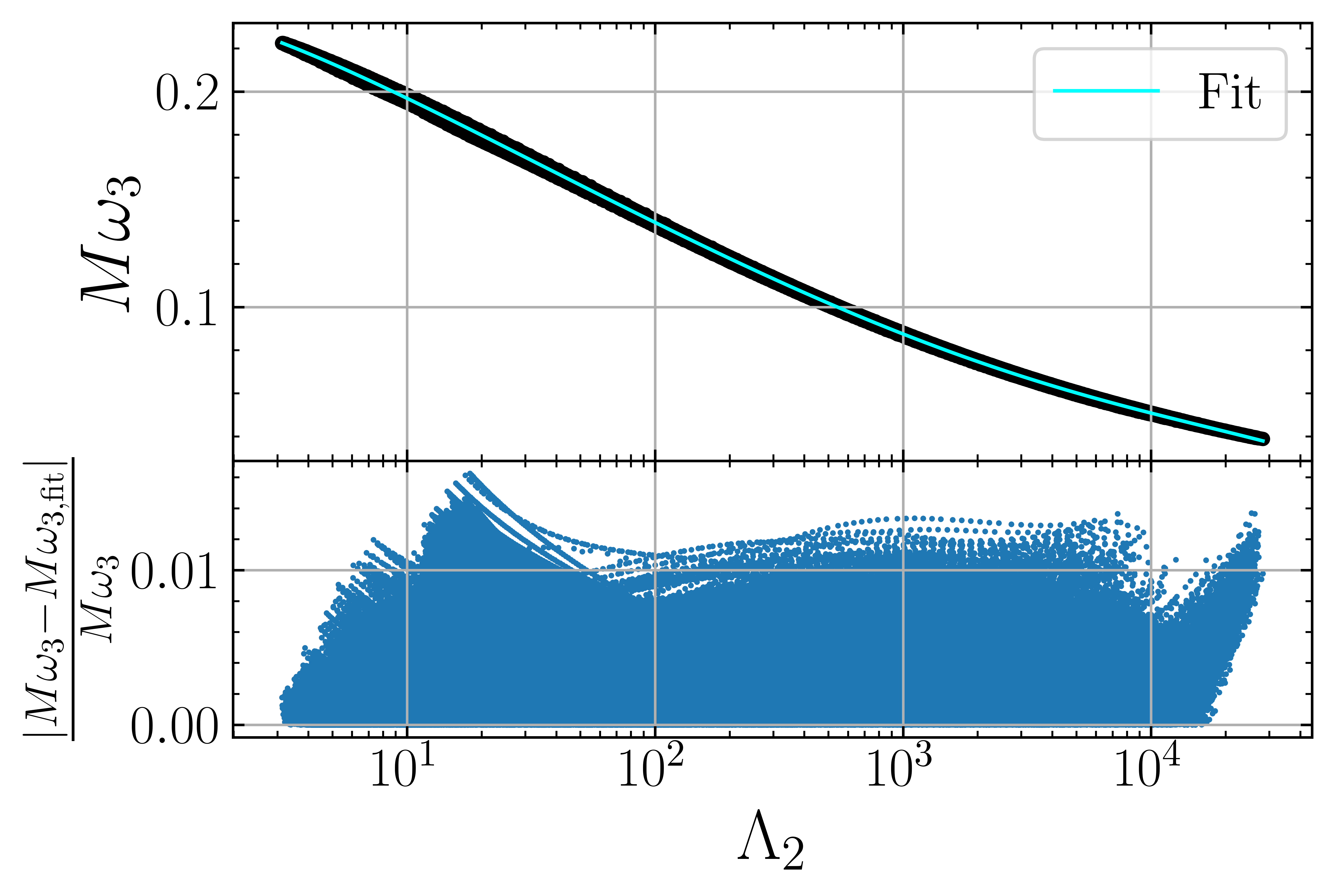}
\caption{}
\label{fig:f_Love_leq3}
\end{subfigure}
 
\begin{subfigure}{\linewidth}
\includegraphics[clip,width=\linewidth]{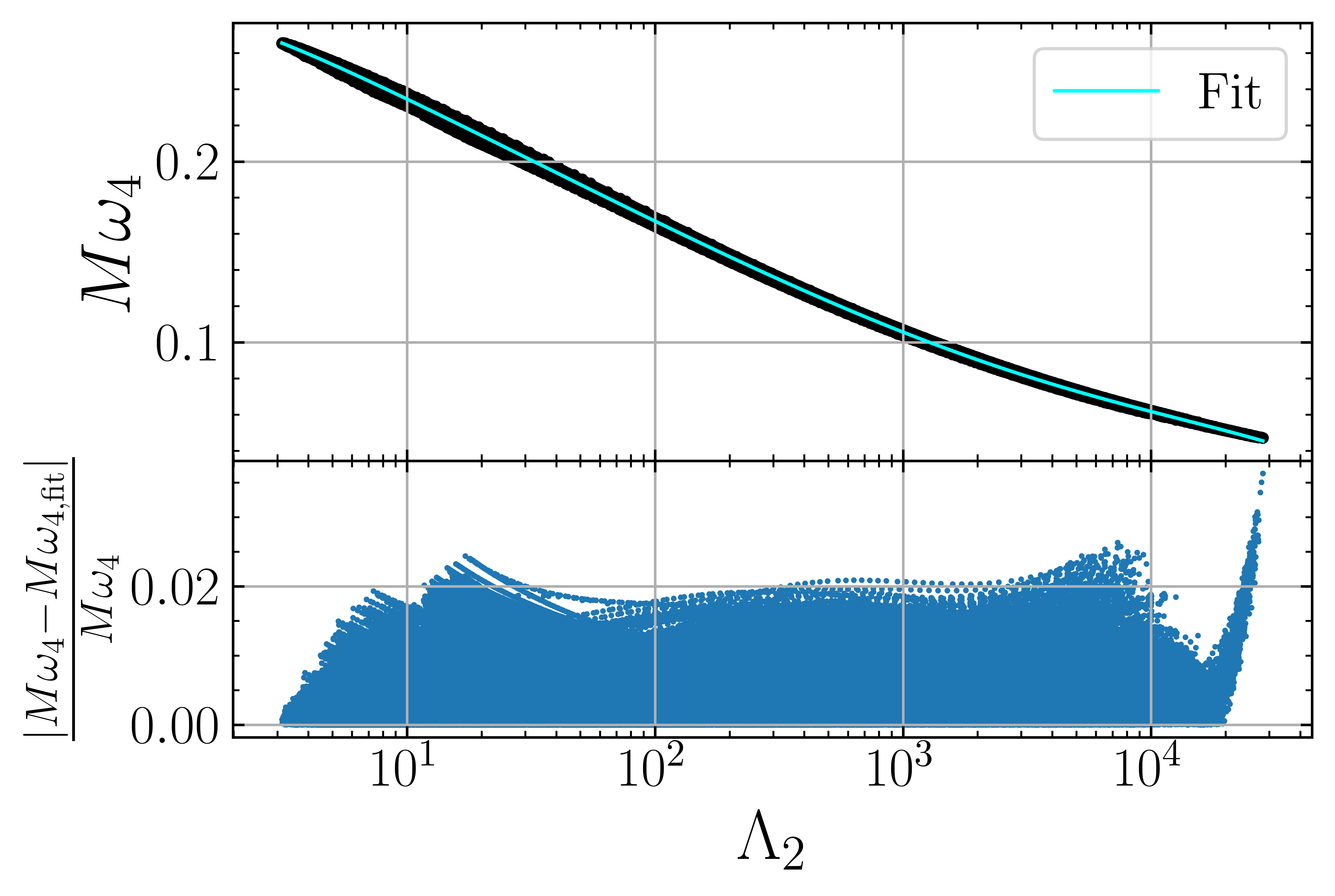}
\caption{}
\label{fig:f_Love_leq4}
\end{subfigure}
\caption{( upper panel ) Variation of mass scaled angular frequency of  f-mode  as a function of $\Lambda_2$ along with the fit relation. (lower panel) Relative error corresponding to fit relation. (a) for quadrupolar ($\ell=2$) f-mode frequency and $\Lambda_2$, (b) for $M\omega_{3}-\Lambda_2$ and  (c) for $M\omega_{4}-\Lambda_2$ UR. }
  \label{fig:meson_fields}

\end{figure}

\subsection{Error Analysis}
We analyze the errors and compare the URs in the range  $\Lambda_2\leq10^4$ as required for GW astronomy. Our $\Lambda_3-\Lambda_2$ relation holds a maximum error of 11\%, with 90\% of the errors are below 7\%. In comparison, the original fit from Yagi~\cite{Yagi} holds a maximum error of 16\% with 90\% of the errors below 10\%, and the updated  $\Lambda_3-\Lambda_2$ UR from Godzieba et al.~\cite{Godzieba2021} holds a maximum error of 23\% with 90\% of the errors below 16\%. On similar lines, the $\Lambda_4-\Lambda_2$ relation developed in this work holds a maximum error of 24\% with 90\% of the errors below 13\%, whereas, the original fit from Yagi holds a maximum error of 35\% with 90\% of the errors below 20\%, and the UR from ~\cite{Godzieba2021} holds a maximum error $\sim$ 44\% with 90\% of the errors below 30\%. For $50\leq \Lambda_2 \leq 10^4$, the URs developed in this work behave quite similarly to the URs developed in~\cite{Godzieba2021} and also in this range, the updated URs, as well as the URs from ~\cite{Godzieba2021} account less error compared to the original fits from Yagi~\cite{Yagi}. However, in the complete range of $\Lambda_2$, our relation have lower error bounds ( see ~\Cref{fig:UR_comparision} in \Cref{sec:UR_compare} or ~\Cref{tab:multipole_Love_fitparameters}).

For $\Lambda_2\leq10^4$, our $\Sigma_2-\Lambda_2$ fit holds a maximum error of 3\% with 90\% of the errors  bellow 1.5\%. The original $\Sigma_2-\Lambda_2$ fit from Yagi~\cite{Yagi_2018} holds a maximum error of  32\% with 90\% of the errors below ~28\%. Although we notice a similar behavior among the  $\Sigma_2-\Lambda_2$ relations from this work and the original fit from Yagi~\cite{Yagi_2018} in the range $3\leq \Lambda_2 \leq 100$, our fit performs much better for $\Lambda_2>100$. Our $\Sigma_3-\Lambda_2$ fit holds a maximum error of 12\% with 90\% of the errors below 5\%. Our $C-\Lambda_2$ relation holds a maximum error $\sim$ 4\% with 90\% of the errors below 2.5\%. The original fit of Maselli et al.~\cite{Maselli} holds a maximum error of ~11\% with 90\% of the errors below 8\%. We notice that our $C-\Lambda_2$ relation always accounts for smaller errors than the other URs in all ranges of $\Lambda_2\leq 10^4$ ( see ~\Cref{fig:UR_comparision} in \Cref{sec:UR_compare} or ~\Cref{tab:C_Love_fitparameters}).

Also, we notice that the universal relation developed in this work involving the quadrupolar f-mode parameter has a maximum error of 0.4\%. In the original work of Chan et al.~\cite{Chan2014}, the universality of $M\omega_{\ell^{\prime}}$ with $\Lambda_{\ell}$ has given for $\ell =\ell^{\prime}$ (though the relation with $\ell \neq \ell^{\prime}$  are plotted) arguing that the relation with $\ell \neq \ell^{\prime}$ can introduce a maximum error. However, in our data set, we notice that the relation $M\omega_3-\Lambda_2$ holds a maximum error  $\sim 1.5\%$ ( in the original fit, the error can reach up to 10\% ) and $M\omega_4-\Lambda_2$  has a maximum error  $\sim 3\%$ which indicates that the relations are tight enough to obtain $\omega_{\ell \geq 2}$ from $\Lambda_2$. The advantages of providing the universal relations for  $M\omega_{\ell \geq 2}$ with $\Lambda_2$ are (i) one can avoid the use of $\Lambda_{\ell \geq 3}-\Lambda_2$ relation which can involve up to 15-20\%  error for further use of $M\omega_{\ell}-\Lambda_{\ell}$ relation (ii) further the relation $M\omega_{\ell \geq 2}-\Lambda_2$ can be directly implemented in the waveform modeling as $\Lambda_2$ is the dominant parameter contributing to the GW phase. We display the f-Love relations for  $M\omega_{2}-\Lambda_{2}$, $M\omega_{3}-\Lambda_{2}$ and $M\omega_{4}-\Lambda_{2}$ in \Cref{fig:f_Love_leq2,fig:f_Love_leq3,fig:f_Love_leq4} respectively. The fit parameters for $M\omega_{\ell}-\Lambda_2$ relations corresponding to  polynomial fit ~\eqref{eqn:tidal_fit} along with relations from Chan et al.~\cite{Chan2014} are tabulated in ~\Cref{tab:f_Love_fitparameters}.

\subsection{GW170817 and Universal relations}
\label{subsec:gw170817}

To see the impact of the multipole Love and f-Love relations, we analyze the event GW170817 within the frequency range 23Hz to 2048Hz with a 128-second signal length.  We consider the inspiral only frequency domain TaylorF2 waveform model with 3.5PN (Post Newtonian) point particle phase, adiabatic tidal effects up to 7.5PN order accounting for the impact from magnetic deformation ($\Sigma_2$), and the octupolar tidal deformability ($\Lambda_3$)~\cite{Henry2020}, and the quadrupolar f-mode dynamical tide at 8PN~\cite{Schmidt2019}. We also consider the spin orbit~\cite{Boh_2013} and spin-spin correction~\cite{Mishra2016} to the waveform model. Additionally, we terminate the waveform at either the contact frequency~\cite{Agathos2015} or the frequency corresponding to stable circular orbit $f_{\rm ISCO}=[6^{3/2}\pi(m_1+m_2)]^{-1}$ , depending on which one is lesser.\\

We perform a  Bayesian parameter estimation of GW170817 strain data~\cite{RICHABBOTT2021}\footnote{\url{https://www.gw-openscience.org/events/GW170817/}} with the noise curve given in ~\cite{noise}, using the dynamic nested sampler \texttt{dynesty} \cite{Dynesty} as implemented in the parameter estimation package \texttt{bilby\_pipe} ~\cite{Bilby_2019}. To speed up the likelihood calculation we use the heterodyned likelihood calculation~\cite{Zackay2018} (also referred to as ``relative binning'') as implemented in \texttt{bilby} \cite{Krishna:2022} ( we use ``relative binning'' for the BNS event GW170817, where later in ~\Cref{subsec:injection_studies} we use the nested sampling  algorithm \texttt{dynesty} as implemented in python package  \texttt{bilby} and not ``relative binning''. ). We use   uniform priors on chirp mass $\mathcal{M}_c \in [1.18,1.21]M_{\odot}$ $\left(\mathcal{M}_c=\frac{(m_1 m_2)^{3/5}}{(m_1+m_2)^{1/5}} \right)$, mass ratio $q\in [0.5,1]$ ($q=m_2/m_1$),  individual  spin magnitudes $|\chi_i| \in [0,0.05]$,  individual quadrupolar  tidal deformability parameter $\Lambda_{2,i}\in [5,5000]$, power law prior on the luminosity distance $d_L\in [1,80]$ Mpc and sine in the binary inclination angle  $\theta_{jn}$ (angle between angular momentum of the binary and line of sight). We fix the right ascension (ra)  and declination (dec) to  -0.4081 radian and ra = 3.446 radian respectively. 
We perform Bayesian  parameter estimation with three different approximants for the waveform
\begin{enumerate}
    \item  TaylorF2 waveform model  with adiabatic tidal effect from quadrupolar tidal deformability  $\Lambda_2$: hereafter referred as TF2,
    \item further we add the adiabatic tidal correction from $\Lambda_3$ and $\Sigma_2$ : this is named as $\rm TF2_{\Lambda_3,\Sigma_2}$ and
    \item then we add the quadrupolar f-mode  dynamical tide ( `\textit{fmtidal}'  phase from ~\cite{Schmidt2019}) and mention it as $\rm TF2_{\Lambda_3,\Sigma_2,f}$. 
\end{enumerate}
To see the impact of universal relations, we consider the previously developed relation from  ~\cite{Yagi,Yagi_2018,Chan2014}  one family of universal relations (mentioned as  `Yagi', if only multipole Love relation is used or `Yagi$+$Chan'  for multipole  Love and f-Love relation in the figures and tables) and the universal relations developed in this work as of another family (mentioned as `This work' in the figures and tables). 

\begin{figure}[htbp]
    \centering
    \includegraphics[width=\linewidth]{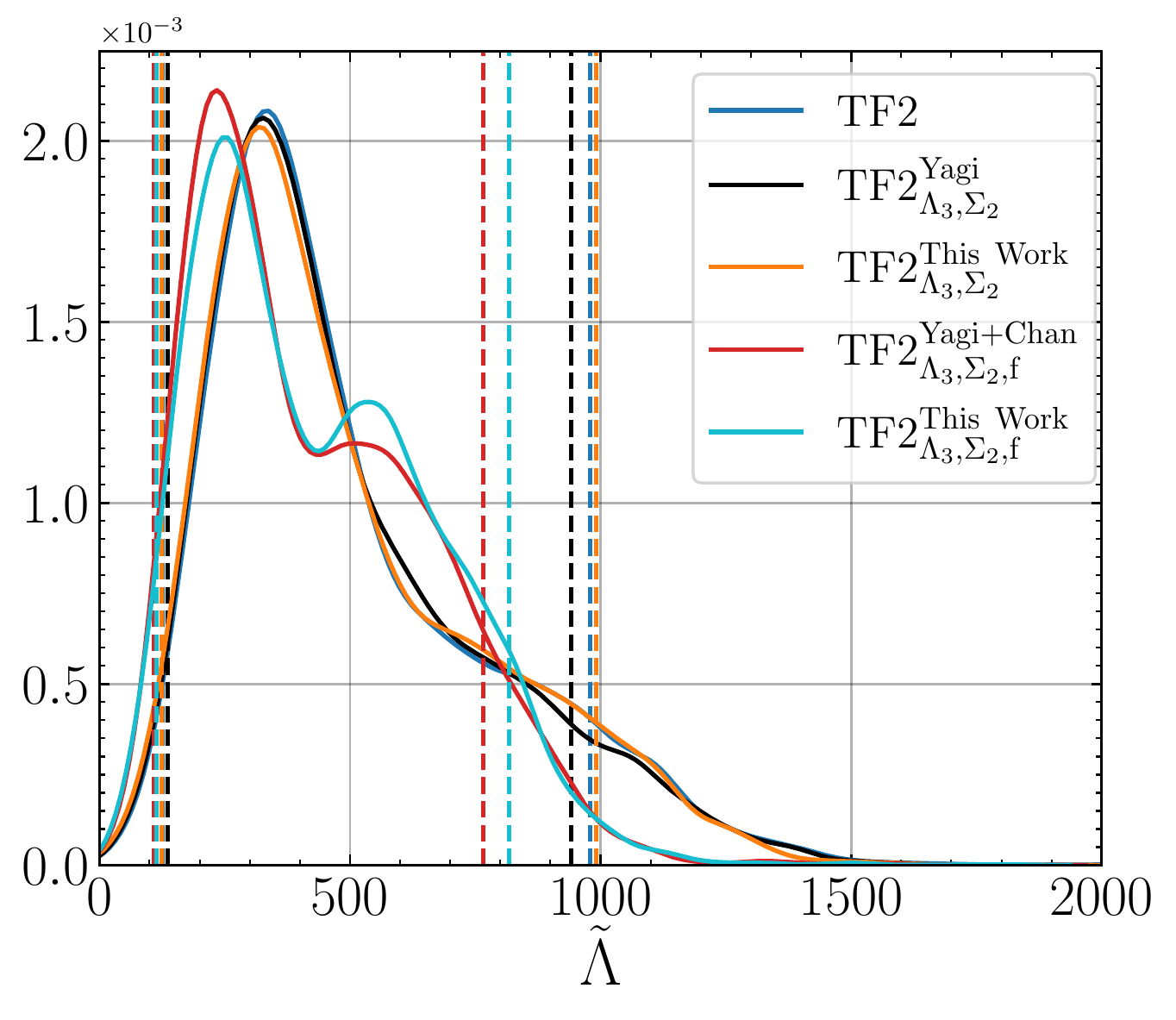}
    \caption{The one-dimensional marginalized  posterior distributions of  $\tilde{\Lambda}$, recovered with the different model corrections and different URs.}
    \label{fig:gw170817_ltilde}
\end{figure}

\begin{figure*}[htbp]
    \centering
    \begin{subfigure}{0.49\textwidth}
      \includegraphics[width=\linewidth]{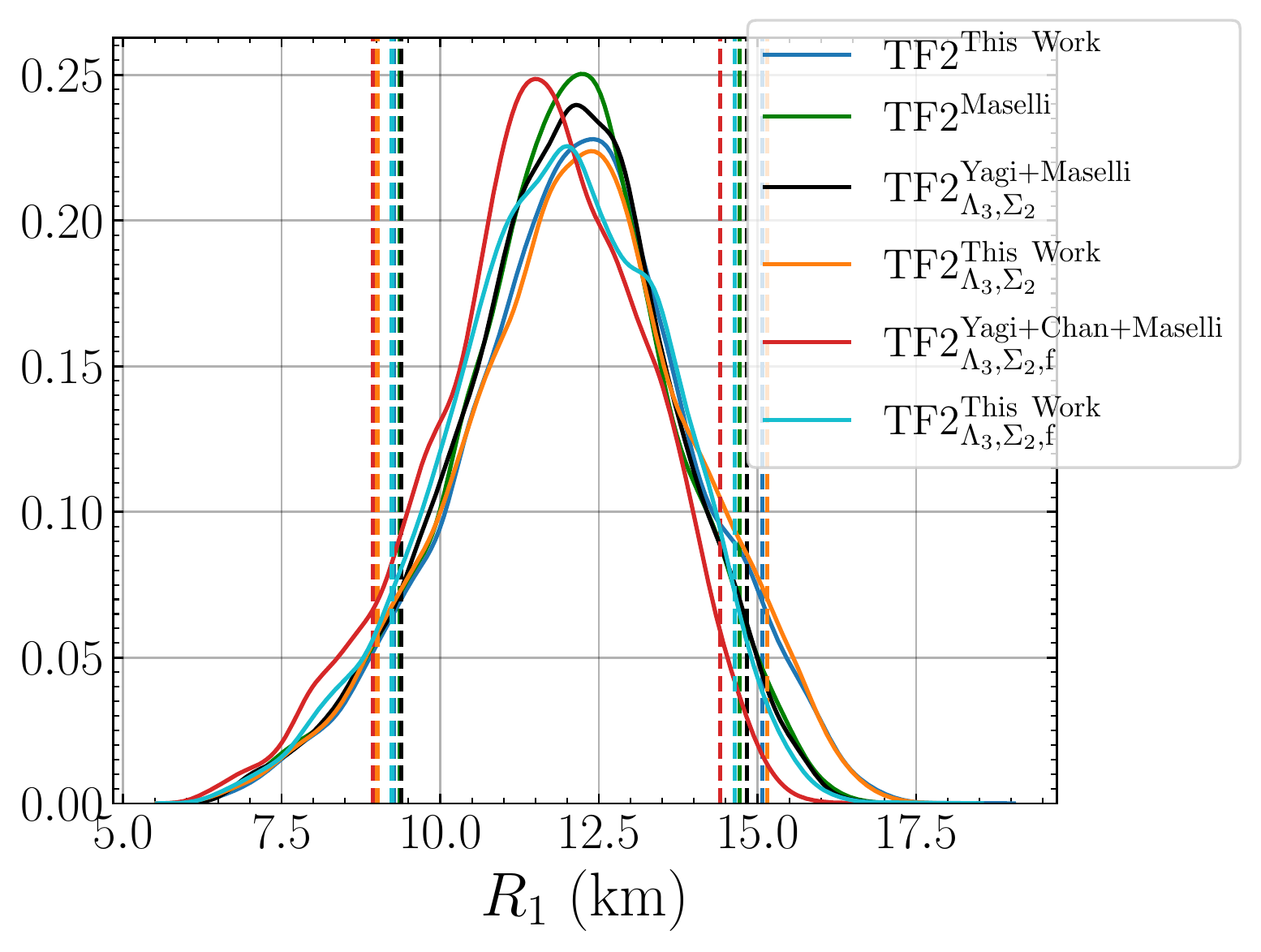}
  \caption{}
  \label{fig:gw170817_r1}
    \end{subfigure}
 \begin{subfigure}{.49\textwidth}
  \centering
  \includegraphics[width=\linewidth]{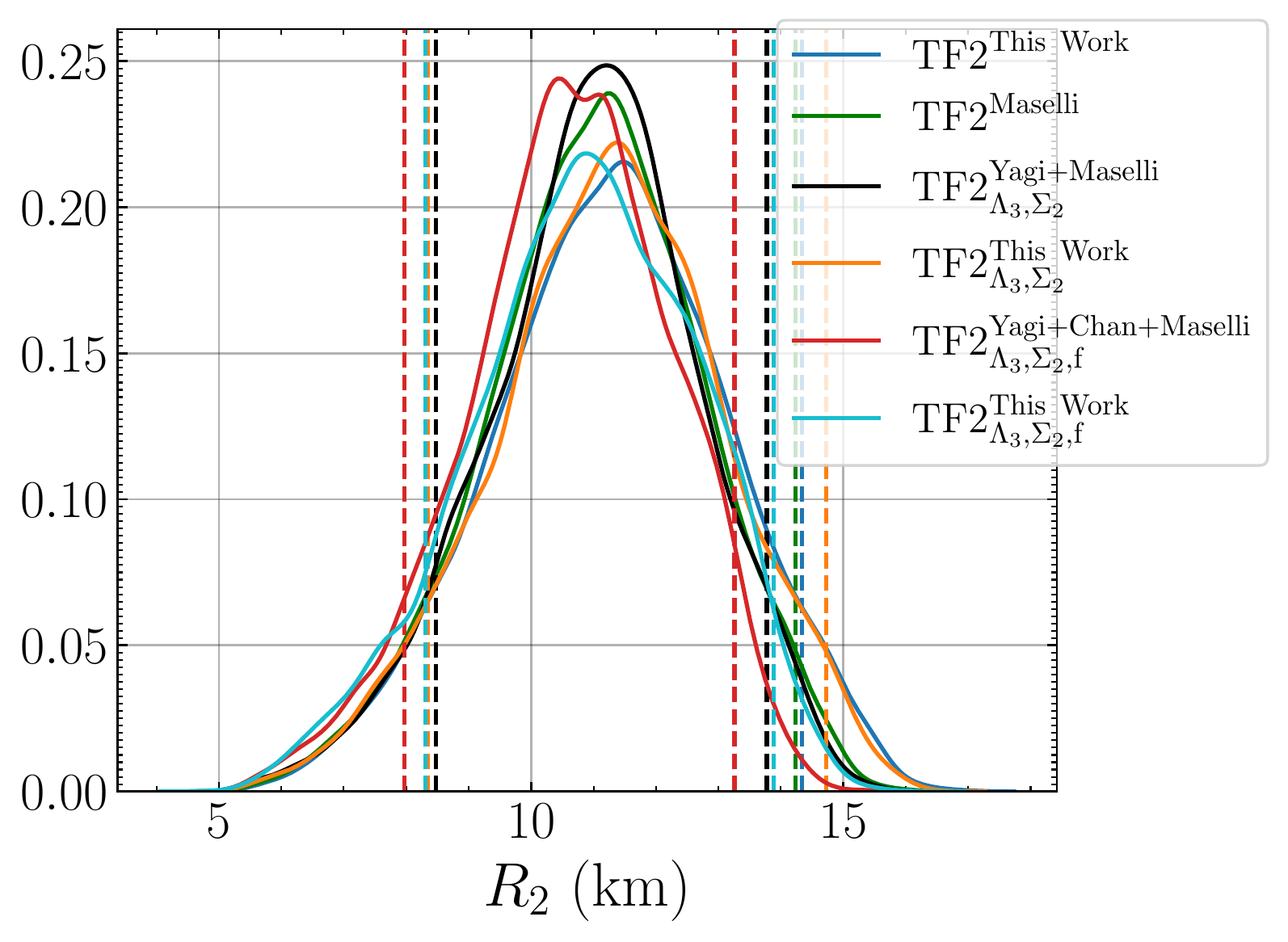}
  \caption{}
  \label{fig:gw170817_r2}
    \end{subfigure}
    \caption{(a) The radius of the primary  component ($m_1$, heavier one) of the binary, estimated through $C-\Lambda_2$ URs from the  tidal parameter ($\Lambda_{2,1}$) and mass distribution ($m_1$). (b) Same as ~\Cref{fig:gw170817_r1} but for the lighter component ($m_2$). Different URs used are labelled in the superscript to the name of waveform  models.}
    \label{fig:gw170817_lr1r2}
\end{figure*}

The important parameters inferred with different waveform model corrections and with different universal relations are tabulated in  ~\Cref{tab:GW170817_pe}. In different considered scenarios, we notice  that the parameters $\mathcal{M}_c$,$q$,$\chi_{\rm eff}$,$d_L$ and $\theta_{jn}$ are recovered all same. For mass ratio ($q$), the upper bound $q\leq1$ is due to the boundary given in the prior, so we only quote the 90\% lower bound on `$q$ .'  The one-dimensional  posterior distribution of the reduced tidal deformability parameter $\tilde{\Lambda}$ (see relation (5) from ~\cite{Wade2014} for definition. ) with different URs and different model corrections are displayed in ~\Cref{fig:gw170817_ltilde}.   From ~\Cref{tab:GW170817_pe} and ~\Cref{fig:gw170817_ltilde}, one can conclude the following regarding the inferred $\tilde{\Lambda}$, the waveform model corrections as well as about the choice of URs:
\begin{itemize}
\item  Addition of tidal phase correction from $\Lambda_3,\ \Sigma_2$ (labelled with  $\rm TF2_{\Lambda_3,\Sigma_2}$ in ~\Cref{fig:gw170817_ltilde}) do not have significant impact on the estimated  $\tilde{\Lambda}$ or the NS radii.  Even the change of multipole Love relations does not affect the estimated $\tilde{\Lambda}$ ( though we notice that the updated URs largely support $\tilde{\Lambda}\leq 300$ reported in ~\cite{AbbottPRX,noise,AbbottPRL119,AbbottPRL121}  compared to the use of previously developed URs.) 
\item Further, adding quadrupolar f-mode dynamical tide  and using the URs from ~\cite{Yagi,Yagi_2018,Chan2014}: i.e, with $\rm TF2_{\Lambda_3,\Sigma_2,f}^{Yagi+Chan}$  model, the median of $\tilde{\Lambda}$ drops by $\sim 11\%$ ( by 45 units) compared to median of $\tilde{\Lambda}$ inferred from $\rm TF2$ model and  drops by $\sim$ 10\% comparing to the median of $\tilde{\Lambda}$ corresponding to $\rm TF2_{\Lambda_3,\Sigma_2}^{Yagi}$ model. 

\item We notice a strong decline of $\sim$22\% ( by $\sim$ 212 units) in the 90\% {highest-probability-density} (HPD) upper bound of $\tilde{\Lambda}$ resulting from $\rm TF2_{\Lambda_3,\Sigma_2,f}^{Yagi+Chan}$  model comparing to the 90\% HPD upper bound of $\tilde{\Lambda}$ resulting from  $\rm TF2$ model. Comparing to the  $\rm TF2_{\Lambda_3,\Sigma_2}^{Yagi}$ model, there is a decline of 19 \% (by $\sim$ 140 units) on the 90\% upper bound of $\tilde{\Lambda}$ after considering the f-mode dynamical tidal effect (i.e, with $\rm TF2_{\Lambda_3,\Sigma_2,f}^{Yagi+Chan}$ model.)~.

\item By choosing the URs developed in this work and considering the f-mode corrected waveform model i.e, with $\rm TF2_{\Lambda_3,\Sigma_2,f}^{This\ Work}$, the median of $\tilde{\Lambda}$ decreases by  $\sim 6\%$ ( by 23 units) compared to median of $\tilde{\Lambda}$ inferred from $\rm TF2$ model and the median drops by $\sim$ 4.5\% compared to the median of $\tilde{\Lambda}$ corresponding to $\rm TF2_{\Lambda_3,\Sigma_2}^{This \ Work}$ model. We notice a decrease  of 16.5\% (by $\sim$ 162 units) in the 90\% HPD upper bound of $\tilde{\Lambda}$ resulting from $\rm TF2_{\Lambda_3,\Sigma_2,f}^{This \ Work}$  model comparing to the 90\% HPD upper bound of $\tilde{\Lambda}$ resulting from  $\rm TF2$ model. Comparing with the  $\rm TF2_{\Lambda_3,\Sigma_2}^{This \ Work}$ model, there is a decline of 17.52 \% (by $\sim$ 173 units) on the 90\% upper bound of $\tilde{\Lambda}$ inferred using  $\rm TF2_{\Lambda_3,\Sigma_2,f}^{This \ Work}$ model. 

\item For $\rm TF2_{\Lambda_3,\Sigma_2,f}$ waveform model, we find that using the URs developed in this work (i.e, with $\rm TF2_{\Lambda_3,\Sigma_2,f}^{This \ Work}$ ) predicts a higher median for $\tilde{\Lambda}$ by 6\% (or by $\sim$ 23 units) than when URs from ~Yagi~\cite{Yagi,Yagi_2018} and Chan et al.~\cite{Chan2014} are used~. Also $\rm TF2_{\Lambda_3,\Sigma_2,f}^{This \ Work}$  predicts a higher  90\% upper bound on $\tilde{\Lambda}$ by 6.7\% ( by 50 units) comparing to $\rm TF2_{\Lambda_3,\Sigma_2,f}^{Yagi+Chan}$.

\end{itemize}
 We estimate radii of the components of BNS event  GW170817 through $C-\Lambda_2$ UR using  the mass and $\Lambda_2$ posterior of the components, i.e,  $(m_1,\Lambda_{2,1}),\textsf{ and }(m_2,\Lambda_{2,2})$.  For the posterior obtained using the previous URs ~\cite{Yagi, Yagi_2018,Chan2014}, we use the $C-\Lambda_2$ UR from Maselli et al.~\cite{Maselli} to infer the NSs radii. However, for the posteriors obtained using the URs from this work, we infer NSs radii using the $C-\Lambda_2$ relation developed in this work. Inferred NS radii of the components of GW170817 resulting from different $C-\Lambda_2$ URs are tabulated in ~\Cref{tab:GW170817_pe}. The inferred NS radius of primary component ($R_1$) and for the secondary component ($R_2$) of GW170817 are displayed in ~\Cref{fig:gw170817_r1,fig:gw170817_r2} respectively. Looking at ~\Cref{fig:gw170817_lr1r2} and ~\Cref{tab:GW170817_pe}, one can conclude the following,
\begin{itemize}
\item  We notice that with the same posterior ($m_1,\ \Lambda_{2,1}$), changing the $C-\Lambda_2$ UR from Maselli et al.~\cite{Maselli} to the $C-\Lambda_2$ UR developed in this work (labeled as $\rm TF2^{This \ Work}$ in ~\Cref{fig:gw170817_r1,fig:gw170817_r2}) the median for $R_1$ increases by $\sim 200\rm m$ (for  secondary component the median of radius $R_2$ increases by 250m due to change of $C-\Lambda_2$ relation), whereas the 90\% upper bound on $R_1$ increase by $\sim$750m using $C-\Lambda_2$ relation from this work comparing to the URs from Maselli et al.~\cite{Maselli} ( for the upper bound of $R_2$ the difference is $\sim$ 600m).

\item Now, using the ($m_1,\Lambda_{2,1}$) or ($m_2,\Lambda_{2,2}$) distribution from $\rm TF2_{\Lambda_3,\Sigma_2}$ model (i.e, with adding the $\Lambda_3$ and $\Sigma_2$ effect) the median of $R_1$ or $R_2$ does not change significantly comparing to the  radii of $\rm TF2$ model.  For waveform model with f-mode corrections and with previously developed universal relations i.e, from ~\cite{Yagi,Yagi_2018,Chan2014,Maselli} ( labeled as $\rm TF2_{\Lambda_3,\Sigma_2,f}^{Yagi+Chan+Maselli}$ in ~\Cref{fig:gw170817_r1,fig:gw170817_r2}  ), predicts a lower median for  $R_1$ by $\sim 400$m  ( or median of  $R_2$ drops by  $\sim 400$m) comparing to the NS radii inferred with $\rm TF2_{\Lambda_3,\Sigma_2,}^{Yagi+Maselli}$ or $\rm TF2^{Maselli}$.

\item Changing the URs to the URs from this work with f-mode dynamical corrections effect, i.e, with $\rm TF2_{\Lambda_3,\Sigma_2,f}^{This \ Work}$ the median of $R_1$ (even for  $R_2$) decreases by 300m comparing to the NS radii informed with $\rm TF2^{This \ Work}$ or $\rm TF2_{\Lambda_3,\Sigma_2}^{This \ Work}$ waveform models. For $\rm TF2_{\Lambda_3,\Sigma_2,f}$ model, comparing the URs developed in this work with that developed in the previous works, we find that updated URs predict a higher median for both $R_1$ and $R_2$ by 300-400m. Inclusion of the f-mode dynamical tidal phase,  the 90\% upper bound on NS radii drops by 400-800m compared to the NS radii estimated by waveform models without the f-mode dynamical phase. 
\end{itemize}
\begin{table*}[ht]
    \centering
    \begin{tabular}{c c c c c c c}
    \hline \hline
    Parameters & TF2 & $\rm TF2_{\Lambda_3,\Sigma_2}^{Yagi}$ & $\rm TF2_{\Lambda_3,\Sigma_2}^{This \ Work}$ & $\rm TF2_{\Lambda_3,\Sigma_2,f}^{Yagi+Chan}$&$\rm TF2_{\Lambda_3,\Sigma_2,f}^{This \ Work}$\\
    \hline
    & & & & & \\
    $\mathcal{M}_c^{\rm source}$  [$M_{\odot}$] & \chirpTFsym & \chirpTFsigYagisym &\chirpTFsigThissym &\chirpTFfmtidalYagiChansym & \chirpTFfmtidalThissym \\
     & & & & & \\
    $q$&\qTFsym&\qTFsigYagisym&\qTFsigThissym&\qTFfmtidalYagiChansym&\qTFfmtidalThissym\\
    & & & & & \\
    $\chi_{\rm eff}$&\chieffTFsym&\chieffTFsigYagisym&\chieffTFsigThissym&\chieffTFfmtidalYagiChansym&\chieffTFfmtidalThissym\\
    & & & & & \\
    $d_L$ [Mpc]&\ldTFsym&\ldTFsigYagisym&\ldTFsigThissym&\ldTFfmtidalYagiChansym&\ldTFfmtidalThissym\\
    & & & & & \\
    $\theta_{jn}$ [rad] &\thetajnTFsym & \thetajnTFsigYagisym & \thetajnTFsigThissym & \thetajnTFfmtidalYagiChansym & \thetajnTFfmtidalThissym\\
    & & & & & \\
    $\tilde{\Lambda}$ (HPD) &  \LtildeTFhpd  & \LtildeTFsigYagihpd  & \LtildeTFsigThishpd  & \LtildeTFfmtidalYagiChanhpd & \LtildeTFfmtidalThishpd \\
    & & & & & \\
    $\tilde{\Lambda}$ (Sym)  & \LtildeTFsym &  \LtildeTFsigYagisym & \LtildeTFsigThissym  & \LtildeTFfmtidalYagiChansym  & \LtildeTFfmtidalThissym\\
    & & & & & \\
    $\delta \tilde{\Lambda}$ & \dlTFsym & \dlTFsigYagisym & \dlTFsigThissym & \dlTFfmtidalYagiChansym & \dlTFfmtidalThissym\\
    & & & & & \\
     
    $R_1 $[km]  (HPD) & \RprimeTFMaselli & \RprimeYagiMasseli & - &\RprimefmYagiChan& -\\
    (Maselli et al.~\cite{Maselli})& & & & & \\
    $R_1 $[km]  (HPD) & \RprimeTFthis & - & \Rprimesigthiswork & - & \Rprimefmthiswork\\
    (This Work)& & & & & \\
    $R_2$[km] (HPD) & \RsecondTFMaselli & \RsecondYagiMasseli & - &\RsecondfmYagiChan&-\\
    (Maselli et al.~\cite{Maselli}) & & & & &\\
    $R_2 $[km]  (HPD) & \RsecondTFthis & - & \Rsecondsigthiswork & - &\Rsecondfmthiswork\\
    (This Work)& & & & & \\
    $\ln \l(\mathcal{B}^M_{TF2}\r)$&0&-0.03&-0.12&-0.63&-0.63\\
    \hline \hline
    \end{tabular}
    \caption{Median and 90\% symmetric credible interval of recovered GW170817  parameters : source frame chirp mass ($\mathcal{M}_c^{\rm source}$  [$M_{\odot}$]), effective spin ($\chi_{\rm eff}$, see Eq. 3 of ~\cite{AbbottPRX} for definition), binary inclination angle ($\theta_{jn}$), $\tilde{\Lambda}$, $\delta \tilde{\Lambda}$. For mass ratio ($q$), the upper bound 1 is restricted by the prior so we note the 90\% symmetric credible lower bound with the upper bound as 1. The different URs used are labeled in the superscript corresponding to each model (e.g., $\rm TF2_{\Lambda_3,\Sigma_2,f}^{Yagi+Chan} $ means that the URs used for $\Lambda_3-\Lambda_2$ and $\Sigma_2-\Lambda_2$ are those given by Yagi ~\cite{Yagi,Yagi_2018} and  for $M \omega_2-\Lambda_2$ the UR relation from ~\cite{Chan2014} is used whereas, models labeled with `This Work' in the superscript use the URs developed in this work). For the reduced tidal parameter $\tilde{\Lambda}$ we also provide the 90\% highest-probability-density (HPD) credible regions. We provide the Bayes factor for different model corrections computed against $\rm TF2$ model.}
    \label{tab:GW170817_pe}
\end{table*}


 \subsection{Injection Studies}
 \label{subsec:injection_studies}
For the BNS event  GW170817, the higher frequency region is mainly dominated by noise and does not show any significant impact regarding the f-mode or due to the change of URs~\cite{Pratten2020, Gamba2022,Godzieba2021}. However, it is anticipated that with the upgraded sensitivity in the future LIGO-VIRGO runs (with A+ configuration ) or even in next-generation detectors (ET or CE), the dynamical f-mode tidal effect or even the change in URs can have a significant impact on the NS properties inferred from BNS events. Recently Pratten et al.~\cite{Pratten2022} concluded that ignoring the dynamical tidal phase can overestimate the NS radii by $\sim$ 10\%. In ~\cite{Pratten2022,Williams2022}, it has been discussed that the impact of the dynamical tide on the inferred tidal parameter depends upon the choice of EoS. Hence, we investigate the impact of the URs in addition to the f-mode dynamical tidal effect (as the literature suggests that this has a dominant impact) and see if our conclusion depends upon the choices of the EoSs.

We consider the  detector configurations for A+ and ET similar to that of ~\cite{Pratten2022}, i.e.,  2 detectors  from LIGO (H1 and L1) and VIRGO with the A+ design sensitivity~\cite{Abbott2020}~\footnote{\url{https://dcc.ligo.org/LIGO-T2000012/public}}, as anticipated for the fifth observing run (O5) and the third generation (3G) Einstein telescope (ET) with ET-D sensitivity~\cite{Hild_2011}~\footnote{\url{https://dcc.ligo.org/LIGO-T1500293/public}}. We inject the simulated BNS waveform using the $\rm TF2_{\Lambda_3,\Sigma_2,f}$ waveform model: i.e, including the tidal correction from additional tidal parameter $\Lambda_3,\Sigma_2$ and f-mode dynamical tide~\cite{Williams2022}. We assume the NSs are nonspinning and the orbits are quasi-circular, i.e., we ignore the individual spins and orbital eccentricity~\cite{Pratten2022}. As in a BNS system the tidal information is mostly contained in the in-spiral phase, we focus on the inspiral waveform only and   truncate the waveform  at a frequency that is the minimum among the  contact frequency ($f_{\rm contact}$) and $f_{\rm ISCO}$, i.e, $f_{\rm max}=\rm{Minimum}$ $(f_{\rm{contact}}$,$f_{\rm ISCO})$. The injected waveform starts at a minimum frequency $ f_{\rm min}=20$ Hz.

We recover the BNS parameters from injected signals with and without the dynamical tide:  a model with adiabatic tidal correction including the additional multipole Love parameters, $\rm TF2_{\Lambda_3,\Sigma_2}$ and another with both adiabatic and dynamical tidal corrections ($\rm TF2_{\Lambda_3,\Sigma_2,f}$). During recovery, the  URs from ~\cite{Yagi, Yagi_2018,Chan2014}  are kept as one family, and the URs developed in this work as of a different family whereas, for $\rm TF2_{\Lambda_3,\Sigma_2}$ model we use the URs developed in this work to recover $\Lambda_3$ and $\Sigma_2$ from $\Lambda_2$. During injection, for $\Lambda_3,\ \Sigma_2$, and $M\omega_2$, we use their actual values corresponding to each EoS and use the UR only while recovering the parameters. We perform the Bayesian parameter estimation using GW data inference package Bilby\_Pipe with nested sampler dynasty as implemented in Bilby~\cite{Bilby_2019}.

To see the biases due to the choices of EoSs, we consider three different EoSs with different stiffness spanning from soft to stiff: Soft-EoS from ~\cite{Hebeler_2013}, an intermediate EoS APR4 ~\cite{APR} and the Stiff-EoS~\cite{Hebeler_2013}. For each EoS, we perform  three different  injection and recover studies: one injection with  source frame chirp mass that of the BNS event  GW170817, $\mathcal{M}_{c,GW170817}^{\rm source}=1.186M_{\odot}$ and other 2 events with $\mathcal{M}_c^{\rm source}=\pm 10\% \mathcal{M}_{c,GW170817}^{\rm source}$, i.e., the considered events have   varying the source frame chirp mass  $\mathcal{M}_c^{\rm source}=\{1.067,1.186,1.304\}M_{\odot}$ and mass ratio $q=0.855$~\cite{Pratten2022}\footnote{In ~\cite{Pratten2022}, the impact of f-mode  dynamical tide is discussed with consideration of a wide range of injections with varying source frame chirp masses in the range $\mathcal{M}_c^{\rm source}=\{0.949,1.067,1.186,1.305,1.423,1.542,1.66\}M_{\odot}$. However, we choose only three configurations to reduce the computational time as we have to investigate the impact of URs simultaneously.}. We include Gaussian noise in our analysis. Following the arguments from ~\cite{Pratten2022}, we focus on the nearby sources ( with luminosity distance $\leq 150$Mpc) and fix the distance and source position during the recovery (which is based upon the assumption that the BNS events can be associated with electromagnetic (EM) counterparts). The priors are uniform in $\mathcal{M}_c$, uniform in symmetric mass ratio $\eta=m_1m_2/(m_1+m_2)^2$ and uniform priors in $\tilde{\Lambda}$ and $\delta \tilde{\Lambda}$. We choose the detector configurations and injection parameters similar to ~\cite{Pratten2022}, and check our numerical method regarding the implementation of the dynamical tide by reproducing results from ~\cite{Pratten2022} for  APR4 EoS and considered mass range.

 We tabulate the important recovered source frame NS parameters for different injected events along with the injected parameters for  Soft-EoS, APR4, and Stiff-EoS in \Cref{tab:Soft_EoS_injections,tab:APR4_injections,tab:Stiff_EoS_injections} respectively. We provide the deviation of the median of recovered $\tilde{\Lambda}$ from the injected value along with a check or cross mark indicating whether or not the injected value of $\tilde{\Lambda}$ is recovered within the symmetric 90\% credible interval during parameter estimation.  

\begin{figure*}[htbp]

\begin{subfigure}{\linewidth}
\includegraphics[clip,width=\linewidth]{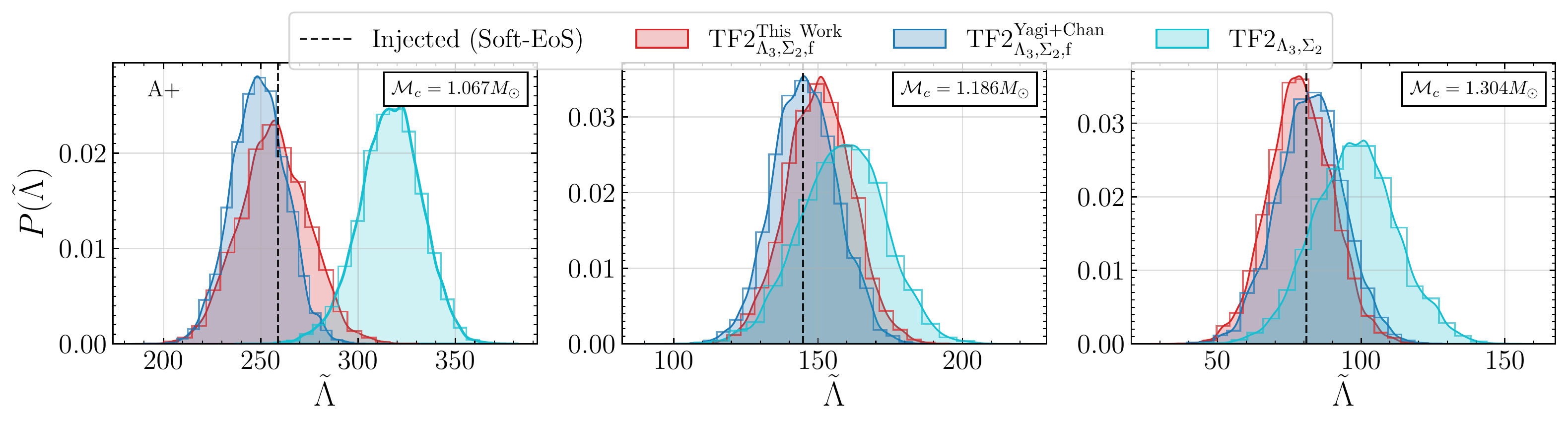}
\caption{}
\label{fig:ltilde_Soft_EoS}
\end{subfigure}

\begin{subfigure}{\linewidth}
\includegraphics[clip,width=\linewidth]{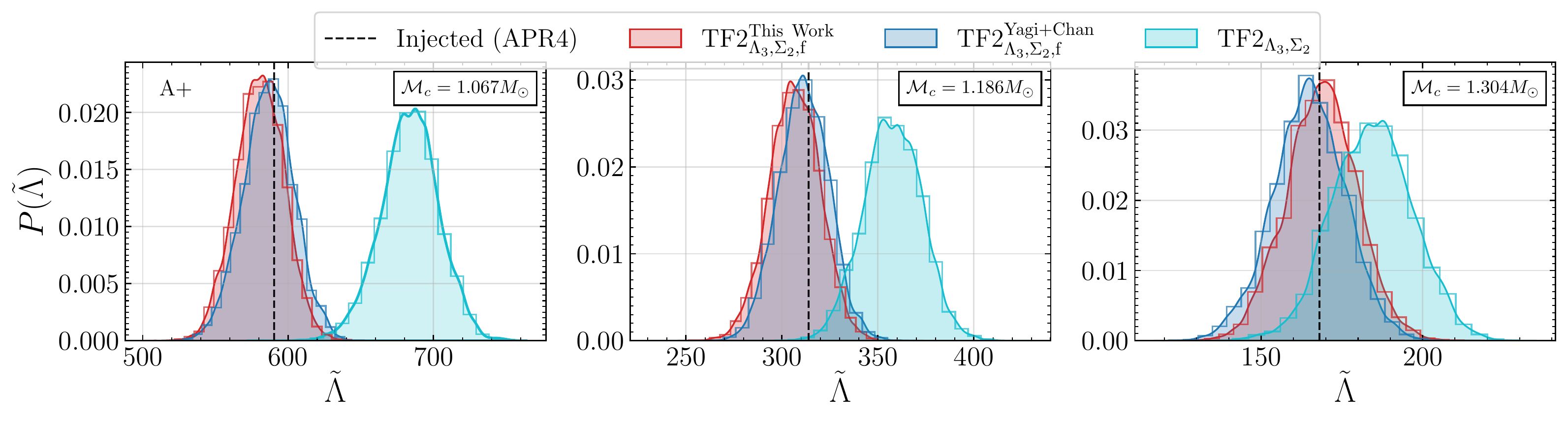}
\caption{}
\label{fig:ltilde_APR4}
\end{subfigure}
 
\begin{subfigure}{\linewidth}
\includegraphics[clip,width=\linewidth]{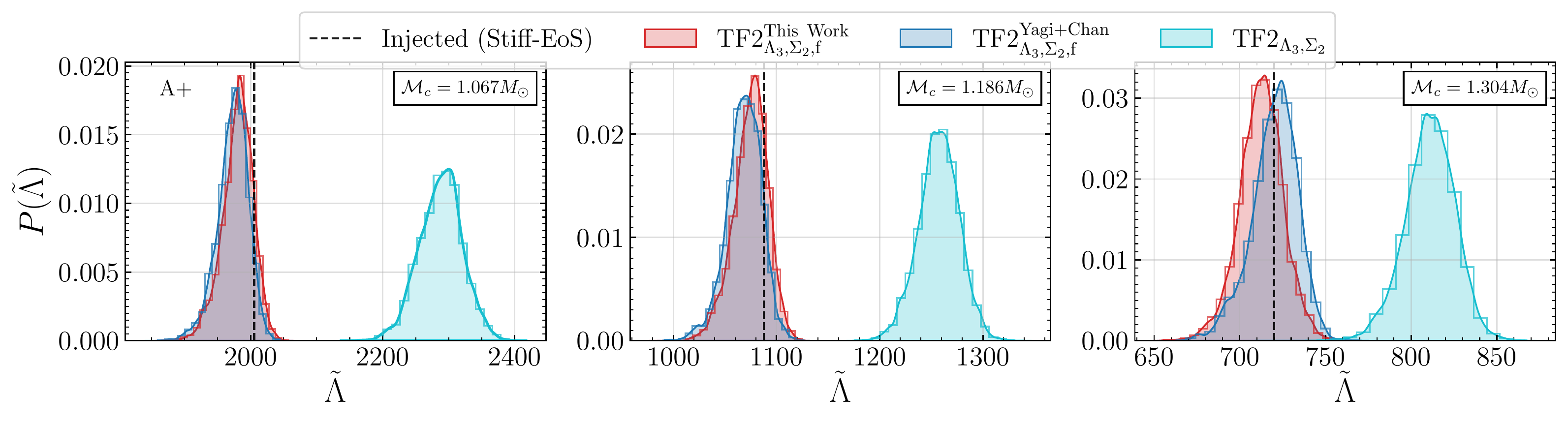}
\caption{}
\label{fig:ltilde_Stiff_EoS}
\end{subfigure}
\caption{Histogram and Posterior distribution of recovered $\tilde{\Lambda}$  for different injection events with A+ detector configuration corresponding to the (a) Soft-EoS, (b) APR4 EoS and  (c)  Stiff-EoS . The injected values are shown with black dashed lines.}
  \label{fig:injection_eos}

\end{figure*}

 For A+ detector configuration, the one-dimensional marginalized posterior distribution  of recovered  reduced tidal parameter $\tilde{\Lambda}$  for different events with  Soft-EoS, APR4 and Stiff-EoS are displayed in ~\Cref{fig:ltilde_Soft_EoS,fig:ltilde_APR4,fig:ltilde_Stiff_EoS} respectively. For injections with Soft-EoS, although the ignorance of f-mode dynamical tide can overestimate the $\tilde{\Lambda}$ up to 10-20\% compare to injected values, {except} for $\mathcal{M}_c=1.067$ the injected value is well recovered within 90\% credible of recovered posterior even ignoring the f-mode dynamical tide. Switching the EoS to the intermediate APR4 and ignoring the dynamical tide in recovery waveform, the injected $\tilde{\Lambda}$ is only recovered for $\mathcal{M}_c=1.304$ (with a 10\% deviation in the median compared to the injected value) within 90\% credible region. For injections with Stiff-EoS, we never recover the injected $\tilde{\Lambda}$ by ignoring the dynamical tide in the recovery waveform, irrespective of the mass ranges considered in this work. However,  all the injection parameters are well recovered by considering the f-mode dynamical tides in the recovery waveform. Additionally, we notice that both the set of URs, i.e., URs from ~\cite{Yagi,Yagi_2018,Chan2014} and the updated URs in this work perform similarly. They both recover the injected parameters very well. The maximum deviation found in the median due to the change of URs is $\sim$5\%.

 \begin{table*}[ht]
    \centering
    \begin{tabular}{c c| c c c| c c c}
    \hline \hline
     & & & A+ & & &ET& \\
     \hline
    Parameters & Injection & $\rm TF2_{\Lambda_3,\Sigma_2}$ & $\rm TF2_{\Lambda_3,\Sigma_2,f}^{\rm Yagi+Chan}$ & $\rm TF2_{\Lambda_3,\Sigma_2,f}^{\rm This\ Work}$   & $\rm TF2_{\Lambda_3,\Sigma_2}$ & $\rm TF2_{\Lambda_3,\Sigma_2,f}^{\rm Yagi+Chan}$ & $\rm TF2_{\Lambda_3,\Sigma_2,f}^{\rm This\ Work}$\\
    \hline
    $\mathcal{M}_c^{\rm source}$  [$M_{\odot}$]& \chirpinjectmconezerosixSoft & \chirpnofmmconezerosixSoft & \chirpfmoldmconezerosixSoft &\chirpfmnewmconezerosixSoft & \chirpnofmmconezerosixETSoft & \chirpfmoldmconezerosixETSoft &\chirpfmnewmconezerosixETSoft \\
    $q$& \qinjectmconezerosixSoft & \qnofmmconezerosixSoft & \qfmoldmconezerosixSoft &\qfmnewmconezerosixSoft & \qnofmmconezerosixETSoft & \qfmoldmconezerosixETSoft &\qfmnewmconezerosixETSoft \\
    $\delta \tilde{\Lambda}$& \dlinjectmconezerosixSoft & \dlnofmmconezerosixSoft & \dlfmoldmconezerosixSoft &\dlfmnewmconezerosixSoft & \dlnofmmconezerosixETSoft & \dlfmoldmconezerosixETSoft &\dlfmnewmconezerosixETSoft \\
    $\tilde{\Lambda}$& \LinjectmconezerosixSoft & \LnofmmconezerosixSoft & \LfmoldmconezerosixSoft &\LfmnewmconezerosixSoft & \LnofmmconezerosixETSoft & \LfmoldmconezerosixETSoft &\LfmnewmconezerosixETSoft \\
    $\Delta M_{\tilde{\Lambda}}$[\%] & --& 22.41 (\xmark)& 3.82(\checkmark) & 1(\checkmark) & 11.6 (\xmark) & 2.3 (\checkmark) & 1.13 (\checkmark)\\
    \hline
    $\mathcal{M}_c^{\rm source}$  [$M_{\odot}$]& \chirpinjectmconeeightSoft & \chirpnofmmconeeightSoft & \chirpfmoldmconeeightSoft &\chirpfmnewmconeeightSoft  & \chirpnofmmconeeightETSoft & \chirpfmoldmconeeightETSoft &\chirpfmnewmconeeightETSoft \\
    $q$& \qinjectmconeeightSoft & \qnofmmconeeightSoft & \qfmoldmconeeightSoft &\qfmnewmconeeightSoft& \qnofmmconeeightETSoft & \qfmoldmconeeightETSoft &\qfmnewmconeeightETSoft \\
    $\delta \tilde{\Lambda}$& \dlinjectmconeeightSoft & \dlnofmmconeeightSoft & \dlfmoldmconeeightSoft &\dlfmnewmconeeightSoft & \dlnofmmconeeightETSoft & \dlfmoldmconeeightETSoft &\dlfmnewmconeeightETSoft \\
    
    $\tilde{\Lambda}$& \LinjectmconeeightSoft & \LnofmmconeeightSoft & \LfmoldmconeeightSoft &\LfmnewmconeeightSoft& \LnofmmconeeightETSoft & \LfmoldmconeeightETSoft &\LfmnewmconeeightETSoft \\
    $\Delta M_{\tilde{\Lambda}}$[\%] & --& 9.77(\checkmark)& 1.1(\checkmark) & 4.25(\checkmark)& 11.15(\xmark)& 2.65(\checkmark) & 1(\checkmark)\\
    \hline
    $\mathcal{M}_c^{\rm source}$  [$M_{\odot}$]& \chirpinjectmconethreeSoft & \chirpnofmmconethreeSoft & \chirpfmoldmconethreeSoft &\chirpfmnewmconethreeSoft  & \chirpnofmmconethreeETSoft & \chirpfmoldmconethreeETSoft &\chirpfmnewmconethreeETSoft\\
    $q$& \qinjectmconethreeSoft & \qnofmmconethreeSoft & \qfmoldmconethreeSoft &\qfmnewmconethreeSoft& \qnofmmconethreeETSoft & \qfmoldmconethreeETSoft &\qfmnewmconethreeETSoft \\
    $\delta \tilde{\Lambda}$& \dlinjectmconethreeSoft & \dlnofmmconethreeSoft & \dlfmoldmconethreeSoft &\dlfmnewmconethreeSoft& \dlnofmmconethreeETSoft & \dlfmoldmconethreeETSoft &\dlfmnewmconethreeETSoft \\
    $\tilde{\Lambda}$& \LinjectmconethreeSoft & \LnofmmconethreeSoft & \LfmoldmconethreeSoft &\LfmnewmconethreeSoft& \LnofmmconethreeETSoft & \LfmoldmconethreeETSoft &\LfmnewmconethreeETSoft \\
    $\Delta M_{\tilde{\Lambda}}$[\%] & --& 21.12(\checkmark)& 1.3 (\checkmark) & 2.3(\checkmark) & 13.69 (\xmark) & 2.3 (\checkmark) & 1.3 (\checkmark)\\
    \hline \hline
    \end{tabular}
    \caption{Median and 90\%  symmetric credible interval of recovered source parameters ($\mathcal{M}_c,\ q, \ \tilde{\Lambda},\ \delta \tilde{\Lambda}$) for different events in $A^+$ and ET configuration. All the events are injected with $\rm TF2_{\Lambda_3,\Sigma_2,f}$ waveform model. The injection parameters correspond to  the Soft-EoS. We also tabulate the deviation of the median of reduced tidal parameter $\tilde{\Lambda}$ from the injection values ($\Delta M_{\tilde{\Lambda}}$). If the injected value of $\tilde{\Lambda}$ is recovered with in the 90\% credible region we give a check mark, else a cross mark is mentioned.}
    \label{tab:Soft_EoS_injections}
\end{table*}

\begin{table*}[ht]
    \centering
    \begin{tabular}{c c| c c c| c c c}
    \hline \hline
     & & & A+ & & &ET& \\
     \hline
    Parameters & Injection & $\rm TF2_{\Lambda_3,\Sigma_2}$ & $\rm TF2_{\Lambda_3,\Sigma_2,f}^{\rm Yagi+Chan}$ & $\rm TF2_{\Lambda_3,\Sigma_2,f}^{\rm This\ Work}$   & $\rm TF2_{\Lambda_3,\Sigma_2}$ & $\rm TF2_{\Lambda_3,\Sigma_2,f}^{\rm Yagi+Chan}$ & $\rm TF2_{\Lambda_3,\Sigma_2,f}^{\rm This\ Work}$\\
    \hline
    $\mathcal{M}_c^{\rm source}$  [$M_{\odot}$]& \chirpinjectmconezerosixAPR & \chirpnofmmconezerosixAPR & \chirpfmoldmconezerosixAPR &\chirpfmnewmconezerosixAPR & \chirpnofmmconezerosixETAPR & \chirpfmoldmconezerosixETAPR &\chirpfmnewmconezerosixETAPR \\
    $q$& \qinjectmconezerosixAPR & \qnofmmconezerosixAPR & \qfmoldmconezerosixAPR &\qfmnewmconezerosixAPR & \qnofmmconezerosixETAPR & \qfmoldmconezerosixETAPR &\qfmnewmconezerosixETAPR \\
    $\delta \tilde{\Lambda}$& \dlinjectmconezerosixAPR & \dlnofmmconezerosixAPR & \dlfmoldmconezerosixAPR &\dlfmnewmconezerosixAPR & \dlnofmmconezerosixETAPR & \dlfmoldmconezerosixETAPR &\dlfmnewmconezerosixETAPR \\
    $\tilde{\Lambda}$& \LinjectmconezerosixAPR & \LnofmmconezerosixAPR & \LfmoldmconezerosixAPR &\LfmnewmconezerosixAPR & \LnofmmconezerosixETAPR & \LfmoldmconezerosixETAPR &\LfmnewmconezerosixETAPR \\
    $\Delta M_{\tilde{\Lambda}}$[\%] & --& 16 (\xmark)& 0.5(\checkmark) & 1.6(\checkmark) & 16 (\xmark)& 1(\checkmark) & 0.03(\checkmark)\\
    \hline
    $\mathcal{M}_c^{\rm source}$  [$M_{\odot}$]& \chirpinjectmconeeightAPR & \chirpnofmmconeeightAPR & \chirpfmoldmconeeightAPR &\chirpfmnewmconeeightAPR  & \chirpnofmmconeeightETAPR & \chirpfmoldmconeeightETAPR &\chirpfmnewmconeeightETAPR \\
    $q$& \qinjectmconeeightAPR & \qnofmmconeeightAPR & \qfmoldmconeeightAPR &\qfmnewmconeeightAPR& \qnofmmconeeightETAPR & \qfmoldmconeeightETAPR &\qfmnewmconeeightETAPR \\
    $\delta \tilde{\Lambda}$& \dlinjectmconeeightAPR & \dlnofmmconeeightAPR & \dlfmoldmconeeightAPR &\dlfmnewmconeeightAPR & \dlnofmmconeeightETAPR & \dlfmoldmconeeightETAPR &\dlfmnewmconeeightETAPR \\
    
    $\tilde{\Lambda}$& \LinjectmconeeightAPR & \LnofmmconeeightAPR & \LfmoldmconeeightAPR &\LfmnewmconeeightAPR& \LnofmmconeeightETAPR & \LfmoldmconeeightETAPR &\LfmnewmconeeightETAPR \\
    $\Delta M_{\tilde{\Lambda}}$[\%] & --& 14(\xmark)& 1(\checkmark) & 2.2(\checkmark) & 15 (\xmark)& 0.4 (\checkmark) & 1.2 (\checkmark)\\
    \hline
    $\mathcal{M}_c^{\rm source}$  [$M_{\odot}$]& \chirpinjectmconethreeAPR & \chirpnofmmconethreeAPR & \chirpfmoldmconethreeAPR &\chirpfmnewmconethreeAPR  & \chirpnofmmconethreeETAPR & \chirpfmoldmconethreeETAPR &\chirpfmnewmconethreeETAPR\\
    $q$& \qinjectmconethreeAPR & \qnofmmconethreeAPR & \qfmoldmconethreeAPR &\qfmnewmconethreeAPR& \qnofmmconethreeETAPR & \qfmoldmconethreeETAPR &\qfmnewmconethreeETAPR \\
    $\delta \tilde{\Lambda}$& \dlinjectmconethreeAPR & \dlnofmmconethreeAPR & \dlfmoldmconethreeAPR &\dlfmnewmconethreeAPR& \dlnofmmconethreeETAPR & \dlfmoldmconethreeETAPR &\dlfmnewmconethreeETAPR \\
    $\tilde{\Lambda}$& \LinjectmconethreeAPR & \LnofmmconethreeAPR & \LfmoldmconethreeAPR &\LfmnewmconethreeAPR& \LnofmmconethreeETAPR & \LfmoldmconethreeETAPR &\LfmnewmconethreeETAPR \\
    $\Delta M_{\tilde{\Lambda}}$[\%] & --& 9.51(\checkmark)& 1.8 (\checkmark) & 0.05(\checkmark)&10.7 (\xmark) & 2.38 (\checkmark) & - (\checkmark)\\
    \hline \hline
    \end{tabular}
    \caption{Same as ~\Cref{tab:Soft_EoS_injections} but the injections are done with APR4 EoS.}
    \label{tab:APR4_injections}
\end{table*}

 \begin{table*}[ht]
    \centering
    \begin{tabular}{c c| c c c| c c c}
    \hline \hline
     & & & A+ & & &ET& \\
     \hline
    Parameters & Injection & $\rm TF2_{\Lambda_3,\Sigma_2}$ & $\rm TF2_{\Lambda_3,\Sigma_2,f}^{\rm Yagi+Chan}$ & $\rm TF2_{\Lambda_3,\Sigma_2,f}^{\rm This\ Work}$   & $\rm TF2_{\Lambda_3,\Sigma_2}$ & $\rm TF2_{\Lambda_3,\Sigma_2,f}^{\rm Yagi+Chan}$ & $\rm TF2_{\Lambda_3,\Sigma_2,f}^{\rm This\ Work}$\\
    \hline
    $\mathcal{M}_c^{\rm source}$  [$M_{\odot}$]& \chirpinjectmconezerosixStiff & \chirpnofmmconezerosixStiff & \chirpfmoldmconezerosixStiff &\chirpfmnewmconezerosixStiff & \chirpnofmmconezerosixETStiff & \chirpfmoldmconezerosixETStiff &\chirpfmnewmconezerosixETStiff \\
    $q$& \qinjectmconezerosixStiff & \qnofmmconezerosixStiff & \qfmoldmconezerosixStiff &\qfmnewmconezerosixStiff & \qnofmmconezerosixETStiff & \qfmoldmconezerosixETStiff &\qfmnewmconezerosixETStiff \\
    $\delta \tilde{\Lambda}$& \dlinjectmconezerosixStiff & \dlnofmmconezerosixStiff & \dlfmoldmconezerosixStiff &\dlfmnewmconezerosixStiff & \dlnofmmconezerosixETStiff & \dlfmoldmconezerosixETStiff &\dlfmnewmconezerosixETStiff \\
    $\tilde{\Lambda}$& \LinjectmconezerosixStiff & \LnofmmconezerosixStiff & \LfmoldmconezerosixStiff &\LfmnewmconezerosixStiff & \LnofmmconezerosixETStiff & \LfmoldmconezerosixETStiff &\LfmnewmconezerosixETStiff \\
    $\Delta M_{\tilde{\Lambda}}$[\%] & --& 14 (\xmark)& 1.5(\checkmark) & 1.1(\checkmark) & 14 (\xmark) & 1.02 (\xmark) & 0.6(\checkmark)\\
    \hline
    $\mathcal{M}_c^{\rm source}$  [$M_{\odot}$]& \chirpinjectmconeeightStiff & \chirpnofmmconeeightStiff & \chirpfmoldmconeeightStiff &\chirpfmnewmconeeightStiff  & \chirpnofmmconeeightETStiff & \chirpfmoldmconeeightETStiff &\chirpfmnewmconeeightETStiff \\
    $q$& \qinjectmconeeightStiff & \qnofmmconeeightStiff & \qfmoldmconeeightStiff &\qfmnewmconeeightStiff& \qnofmmconeeightETStiff & \qfmoldmconeeightETStiff &\qfmnewmconeeightETStiff \\
    $\delta \tilde{\Lambda}$& \dlinjectmconeeightStiff & \dlnofmmconeeightStiff & \dlfmoldmconeeightStiff &\dlfmnewmconeeightStiff & \dlnofmmconeeightETStiff & \dlfmoldmconeeightETStiff &\dlfmnewmconeeightETStiff \\
    
    $\tilde{\Lambda}$& \LinjectmconeeightStiff & \LnofmmconeeightStiff & \LfmoldmconeeightStiff &\LfmnewmconeeightStiff& \LnofmmconeeightETStiff & \LfmoldmconeeightETStiff &\LfmnewmconeeightETStiff \\
    $\Delta M_{\tilde{\Lambda}}$[\%] & --& 16(\xmark)& 1.7(\checkmark) & 1(\checkmark) & 16 (\xmark) & 1.1 (\xmark) & 0.7 (\checkmark)\\
    \hline
    $\mathcal{M}_c^{\rm source}$  [$M_{\odot}$]& \chirpinjectmconethreeStiff & \chirpnofmmconethreeStiff & \chirpfmoldmconethreeStiff &\chirpfmnewmconethreeStiff  & \chirpnofmmconethreeETStiff & \chirpfmoldmconethreeETStiff &\chirpfmnewmconethreeETStiff\\
    $q$& \qinjectmconethreeStiff & \qnofmmconethreeStiff & \qfmoldmconethreeStiff &\qfmnewmconethreeStiff& \qnofmmconethreeETStiff & \qfmoldmconethreeETStiff &\qfmnewmconethreeETStiff \\
    $\delta \tilde{\Lambda}$& \dlinjectmconethreeStiff & \dlnofmmconethreeStiff & \dlfmoldmconethreeStiff &\dlfmnewmconethreeStiff& \dlnofmmconethreeETStiff & \dlfmoldmconethreeETStiff &\dlfmnewmconethreeETStiff \\
    $\tilde{\Lambda}$& \LinjectmconethreeStiff & \LnofmmconethreeStiff & \LfmoldmconethreeStiff &\LfmnewmconethreeStiff& \LnofmmconethreeETStiff & \LfmoldmconethreeETStiff &\LfmnewmconethreeETStiff \\
    $\Delta M_{\tilde{\Lambda}}$[\%] & --& 12.6(\xmark)& - (\checkmark) & 1 (\checkmark) & 15 (\xmark) & 1.13 (\checkmark) & 0.7 (\checkmark)\\
    \hline \hline
    \end{tabular}
    \caption{Same as ~\Cref{tab:Soft_EoS_injections} but the injections are done with Stiff EoS.}
    \label{tab:Stiff_EoS_injections}
\end{table*}
 
For injection and recover studies with ET, we display the distributions of recovered $\tilde{\Lambda}$ for different considered scenarios in ~\Cref{fig:injection_eos_ET} of ~\Cref{sec:ltilde_ET}. It is interesting to conclude that with the ET, we do not recover the injected value of $\tilde{\Lambda}$ within the 90\% credible interval by ignoring the dynamical tide in the recovery waveform irrespective of the considered masses and EoSs. This indicates that the f-mode dynamical tide has much higher impact on 3G detectors compared to A+ configuration (this is expected as the f-mode dynamical effect dominates at high frequency and   3G detectors are more sensitive in high-frequency regimes compared to A+ configurations ). Though the choice of URs changes the median of recovered posteriors of $\tilde{\Lambda}$ only by $\leq$5\%, the distributions get slightly more affected with ET compared to A+ configuration depending upon the choice of URs. For the Stiff-EoS~\cite{Hebeler_2013}, we barely recover the injected $\tilde{\Lambda}$ using the previous URs from ~\cite{Yagi,Yagi_2018,Chan2014}. The biases in the posteriors due to the choices of URs increase with the increase of the mass or the stiffness of the EoS (mostly with the injections corresponding to lower $\tilde{\Lambda}$ value). Also, use of updated  URs recovers the injected $\tilde{\Lambda}$ more efficiently compared to existing URs (e.g, see  ~\Cref{fig:injection_eos_ET}) , which may be due to the better performance of our URs in the lower $\Lambda_2$ values compared to the existing URs~\cite{Yagi,Yagi_2018,Chan2014}.

Our results regarding the dominant effect of the f-mode dynamical tide at low mass NS and the increasing impact with increasing the stiffness of the EoS are consistent with the results from ~\cite{Pratten2022}. This can be explained in the following way: the dynamical tidal phase is $\propto \Lambda_2/(M\omega_2)^2$~\cite{Schmidt2019,Pratten2022} and a Stiff-EoS corresponds to larger $\Lambda_2$ and lower $M\omega_2$~\cite{Pradhan2022} indicating a higher dynamical tidal effect ( even for a fixed EoS with increasing mass, $\Lambda_2$ decreases hence $M\omega_2$ increases (see,~\Cref{fig:f_Love_leq2}) with  decreasing the impact of dynamical tide ).

 \section{Conclusions}
 \label{sec:conclusion}
 We update the multipole Love universal relation and f-Love universal relations by considering a wide range of EoSs originating from different physical motivations and applying astrophysical constraints. We consider the uncertainties in the nuclear and hyper-nuclear parameters by considering the state-of-art relativistic mean field model to describe the NS matter. In addition, we consider 15 other realistic EoSs, including three hybrid-quark matter EoSs and a sample of polytropic EoSs described by the spectral decomposition method. We provide the updated URs for electric type tidal parameter  $\Lambda_{\ell}-\Lambda_2$ for $\ell \leq 4$ ( see \Cref{tab:multipole_Love_fitparameters}) and  for magnetic tidal deformation $\Sigma_{\ell}-\Lambda_3$ (for $\ell \leq 3$, see ~\Cref{tab:multipole_Love_fitparameters} )  as required to connect the higher order tidal parameter with the quadrupolar tidal parameter $\Lambda_2$ related to the correction in GW models. The original fits for the tidal parameters can be found in ~\cite{Yagi,Yagi_2018}, note that in ~\cite{Yagi_2018} the UR is only given for $\Sigma_2-\Lambda_2$. Over the range $2\leq \Lambda_2 \leq 10^4$, our multipole Love URs $\Lambda_3-\Lambda_2$, $\Lambda_4-\Lambda_2$, $\Sigma_2-\Lambda_2$ and $\Sigma_3-\Lambda_2$ hold maximum error of 11\%, 24\%, 3\%  and 12\% respectively.
 
 Further, unlike the original work of Chan et al.~\cite{Chan2014}, where the URs for $M\omega_{\ell^{\prime}}-\Lambda_{\ell}$ are given only for $\ell=\ell^{\prime}$ (arguing that $\ell \neq \ell^{\prime}$ can introduce large error) we provide the URs for $M\omega_{\ell}$ with $\Lambda_2$ irrespective that $\ell$=2 or not (see ~\Cref{tab:f_Love_fitparameters}). Our updated URs $M\omega_2-\Lambda_2$, $M\omega_3-\Lambda_2$ and $M\omega_4-\Lambda_2$ fits have a maximum error of 0.4\%, 1.5\% and 3\% respectively (in the original fits from Chan et al.~\cite{Chan2014} the maximum errors are 1.5\%, 10\%, 15\%). Furthermore, we update the $C-\Lambda_2$ UR useful for estimating the NS radius from the mass and quadrupolar tidal parameter posterior obtained from the GW observational events involving NSs. Our updated $C-\Lambda_2$ relation holds a maximum error $\leq 5\%$  ( see ~\Cref{tab:C_Love_fitparameters}).

To see the effect of URs, we analyze the BNS event GW170817 with inspiral only frequency domain waveform model TaylorF2  and also adding different waveform model corrections or changing the URs. An investigation of summarized points of ~\Cref{subsec:gw170817} leads one to conclude the following: (1) Including the higher order $\Lambda_3$ and magnetic $\Sigma_2$ tidal effect to the GW phase, have no significant effect on the estimated tidal parameter, (2) by the change of URs for multipole Love relation (i.e., for $\Lambda_3$ and $\Sigma_2$) with  $\rm TF2_{\Lambda_3,\Sigma_2}$ waveform model from the URs of Yagi~\cite{Yagi} to the URs developed in this work, we do not notice any significant change in  $\tilde{\Lambda}$ (although our URs favor a lower value for the lower bound of $\tilde{\Lambda}$ and a higher value for the 90\% upper bound). (3) Additionally,  the inclusion of f-mode dynamical tidal correction in the waveform model puts a tighter constraint on the range of $\tilde{\Lambda}$ by decreasing the 90\% upper bound by 16-22\% compared to $\tilde{\Lambda}$ estimated with waveform model phase considering only adiabatic tidal phase. Also, the inclusion of f-mode dynamical tide favors a lower median for  $\tilde{\Lambda}$ (by 6-10\%) compared to other waveform model correction scenarios. Including  f-mode dynamical tidal phase, the updated URs predict a higher  median and higher 90\% upper bound on $\tilde{\Lambda}$ (by $\sim$6\%) compared to the  URs from  ~\cite{Yagi,Yagi_2018,Chan2014}.

We estimate the NS radii for the components of GW170817 using the estimated  mass ($m$) and $\Lambda_2$ posteriors through $C-\Lambda_2$ relations. Our updated $C-\Lambda_2$ UR predicts a higher median  (also  higher upper bound) for  $R_1$ (or $R_2$) by 200-250m compared to use of $C-\Lambda_2$ relation from Masellli at al.~\cite{Maselli}. Further adding the effect from $\Lambda_3$ and $\Sigma_2$ to the waveform phase or changing the multipole Love relation, we did not notice any significant change in the inferred NS radii. However considering the dynamical f-mode tidal phase, the median of both $R_1$ and $R_2$ drops by $\sim$300-400m compared with the radii estimated with waveform without dynamical f-mode phase (the 90\% upper bounds on radii  drop by 400-800m depending upon the choice of URs ). Our updated URs predicts higher radii with  $\rm TF2_{\Lambda_3,\Sigma_2,f}$ compared to the  URs from  ~\cite{Yagi,Yagi_2018,Chan2014,Maselli}.  We also notice that adding f-mode dynamical tidal phase largely supports that $\tilde{\Lambda}\leq 300$ as reported in ~\cite{AbbottPRX,noise} with waveform models with numerical relativity (NR)-tuned tidal effects. {Although the medians of NS radii differ by 2\%-3\%, the upper bound on radii differ by 5\%-7\% after considering  f-mode dynamical tidal correction, which is significant considering the fact that the difference arises from a single event.}  
 
In particular, recent gravitational waveform modelling developments include the dynamical tidal correction due to the excitation of NS f-modes ~\cite{Schmidt2019,Kuan2022,Pnigouras2022,Gamba2022}. Recently in ~\cite{Gamba2022}, by considering the state of the art effective one body (EOB) model TEOBResumS, it has been shown that due to the noise dominance in the higher frequency range of GW170817, both waveform model with and without f-mode dynamics behave similarly. Comparing the Bayes factors from ~\Cref{tab:GW170817_pe} for different models, we also notice that the waveform models with f-mode dynamical correction behave similarly to the waveform models without considering f-mode dynamical tides. The NS properties reported in this work for BNS event GW170817 can be further enhanced by adding the spin-quadrupole~\cite{Nagar2019}, spin-tidal  effect  \cite{Abdelsalhin2018} and spin correction to the f-mode dynamical phase or even by adding eccentricity.

We perform injection and recovery studies with $\rm A+$ sensitivity and ET by considering three different EoSs with different stiffness. In agreement with ~\cite{Pratten2022}, we notice that ignoring the f-mode dynamical tide can overestimate the median  $\tilde{\Lambda}$ by $\sim 10-20\%$  depending upon the EoS and the dynamical tide has a significant impact for low mass NSs, which further enhanced with increasing the stiffness of the EoS. Additionally, here we show that the effect of re-calibrating the URs is subdominant to neglecting f-mode dynamical tides. Although the median of the recovered tidal parameters with a different set of URs differs only by $\sim 5\%$, the distribution of $\tilde{\Lambda}$ gets slightly more affected with ET compared to A+ configuration. In ET, the injected  $\tilde{\Lambda}$ is recovered more efficiently by using the updated URs of this work compared to existing URs. Although the difference between URs  is not statistically discernible on a per-event basis, it is potentially important while combining constraints from many events.

Our conclusions regarding  the injection and recovery studies  may change by considering the additional effect resulting from  spin   and also with the spin correction to the f-mode dynamical phase. Although we ignore the spin and eccentricity, it was suggested that spin and eccentricity further enhance the excitation of f-modes in binary ~\cite{Steinhoff2021}. However, the corrections to the f-mode dynamical tide due to  spin and eccentricity are still matter of investigation and recent efforts are going on this direction ~\cite{Chaurasia2018,Steinhoff2021,Kuan2022,Pnigouras2022}.

The detection of f-mode characteristics  or the post-merger peak frequency $\rm f_{peak}$ can  be used to infer the NS EoS, presence of phase transition or even  the  NS interior composition ~\cite{Blacker2020,Weih2020,Pradhan2021,kumar2021,das2021,Pradhan2022,Hong_2022,Mu2022,Vijaykumar2022,Shamim2022}.Although the detection of f-modes needs the third generation detector sensitivity ~\cite{Williams2022,Pratten2020}, the impact is significant for next observing runs~\cite{Pratten2022}. The detection of  $\rm f_{peak}$ becomes more likely in the next observing runs or even with a third generation detector if the merger happens near a supermassive black hole  ~\cite{Vijaykumar2022}. In Pradhan et al.~\cite{Pradhan2022}, we have shown that the plane of f-mode frequency and  $\Lambda_2$ obtained from NSs in binary can put insights regarding the presence of hyperons in the NS interior. Detection of f-mode frequency can further be used to constrain nuclear  parameters~\cite{Pradhan2022,Sotani2021,Tonetto2021,Kunjipurayil2022}. An important breakthrough may also come with the launch of an optimised GW detector to study post-merger nuclear physics in the frequency range 2–4 kHz, as proposed by the ARC Centre of Excellence for Gravitational Wave Discovery in Australia:   Neutron Star Extreme Matter Observatory  (NEMO)~\cite{nemo}.

\section{Acknowledgements}
We thank Patricia Schmidt, Geraint Pratten and Rahul Kashyap for  a careful reading of our manuscript. This material is based upon work supported by NSF's LIGO Laboratory which is a major facility fully funded by the National Science Foundation. B.K.P thanks Tathagata Ghosh, Rossella Gamba, Kanchan Soni and  Apratim Ganguly for the  useful discussions regarding waveform modelling and the  package \texttt{bilby} . B.K.P is also thankful to Dhruv Pathak, Suprovo Ghosh and Swarnim Shirke for their useful comments regarding EoSs and the writing of the manuscript. {The authors thank the anonymous referee for the appreciation of this work and thoughtful comments}. The authors gratefully acknowledge the use of  Sarathi cluster at IUCAA accessed  through the LIGO-Virgo-Kagra Collaboration. B.K.P acknowledge the  IUCAA HPC computing facility for the computational/numerical work. AV is supported by the Department of Atomic Energy, Government of India, under Project No. RTI4001. AV is also supported by a Fulbright Program grant under the Fulbright-Nehru Doctoral Research Fellowship, sponsored by the Bureau of Educational and Cultural Affairs of the United States Department of State and administered by the Institute of International Education and the United States-India Educational Foundation. We acknowledge  International Centre for Theoretical Sciences (ICTS) for participating in the program - ICTS Summer School on Gravitational-Wave Astronomy (code: ICTS/GWS-2022/5). This work makes use of \texttt{NumPy} \cite{vanderWalt:2011bqk}, \texttt{SciPy} \cite{Virtanen:2019joe}, \texttt{astropy} \cite{2013A&A...558A..33A, 2018AJ....156..123A}, \texttt{Matplotlib} \cite{Hunter:2007}, \texttt{jupyter} \cite{jupyter}, \texttt{pandas} \cite{mckinney-proc-scipy-2010} \texttt{dynesty} \cite{Dynesty}, \texttt{bilby} \cite{Bilby_2019} and \texttt{PESummary} \cite{Hoy:2020vys} software packages. 
\appendix
\section{Comparing URs}
\begin{figure*}[htbp]
    \begin{subfigure}{.45\textwidth}
  \centering
  \includegraphics[width=\linewidth]{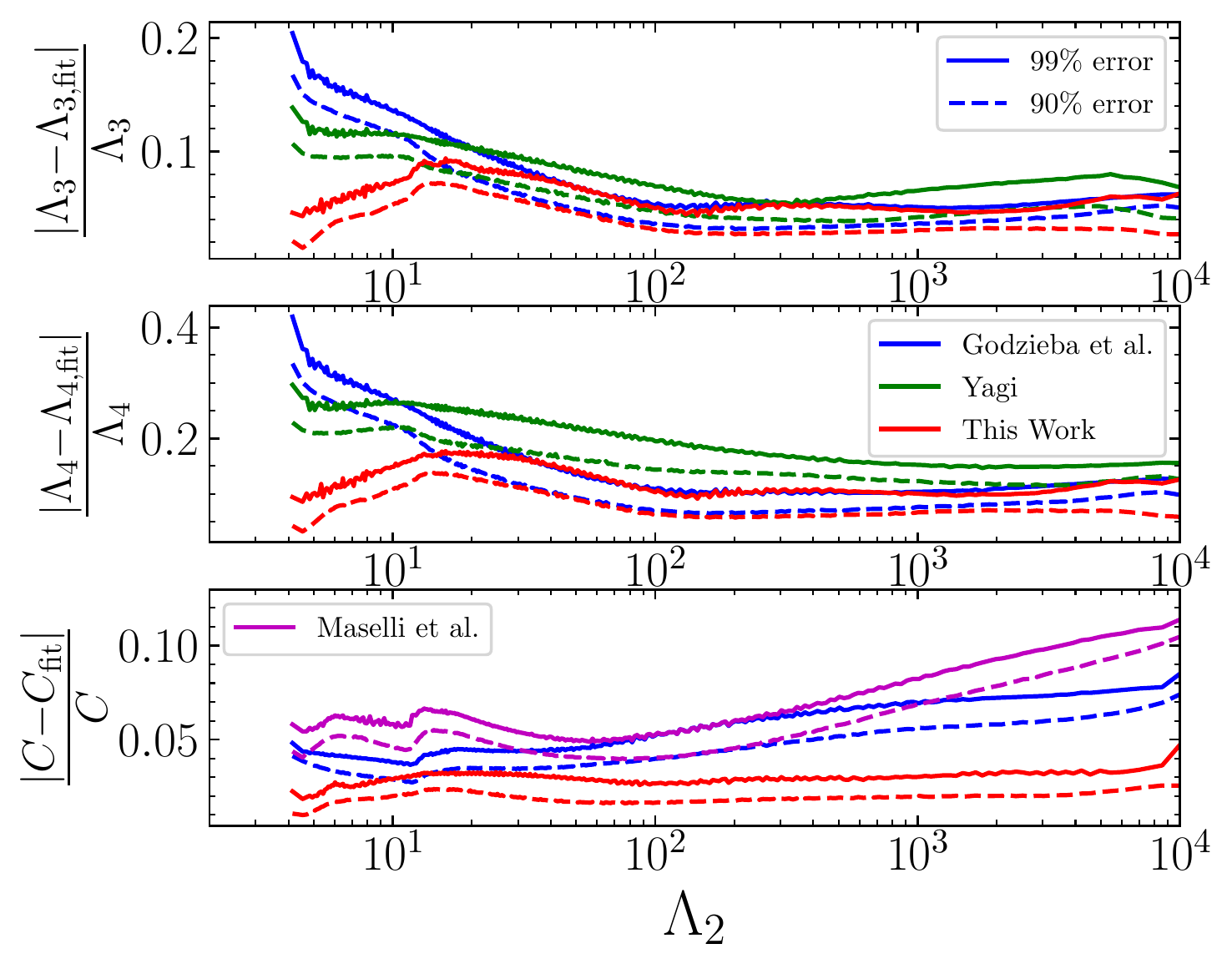}
  \caption{}
    \end{subfigure}
    \begin{subfigure}{0.45\textwidth}
      \includegraphics[width=\linewidth]{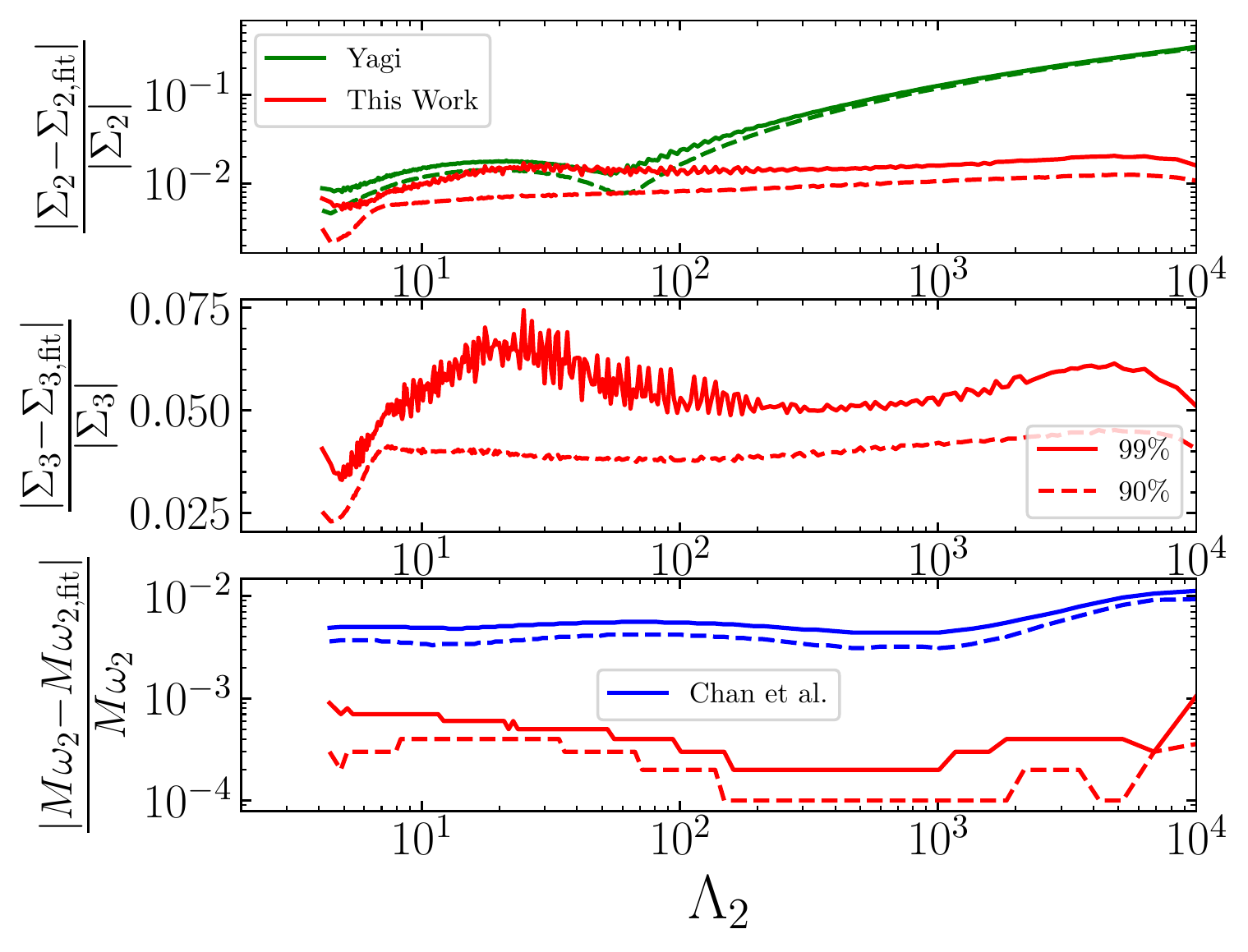}
  \caption{}
    \end{subfigure}
    \caption{Relative errors for  higher order  electric tidal parameters ($\Lambda_3$, $\Lambda_4$), magnetic tidal parameter $\Sigma_2$, stellar compactness ($C$) and mass scaled f-mode angular frequency ($M\omega$) resulting from different URs in the range $\Lambda_2\leq10^4$. Solid line (dashed ) represents the upper limit on the 99\%(90\%) of the errors. Different colored line represents different URs as labelled in the figures. If necessary, for better visibility the y-axis is scaled with logarithmic scale.}
    \label{fig:UR_comparision}
\end{figure*}

 We compare different URs by comparing the relative errors corresponding to different URs. The relative errors over the range of $\Lambda_2\leq 10^4$ for  $\Lambda_3$, $\Lambda_4$,  $\Sigma_2$, $\Sigma_3$, $C$ and $M\omega_2$ resulting from different URs are displayed in ~\Cref{fig:UR_comparision}.
\label{sec:UR_compare}
\section{Distribution of $\tilde{\Lambda}$ of simulated events in ET configuration}
\label{sec:ltilde_ET}
\begin{figure*}[htbp]

\begin{subfigure}{\linewidth}
\includegraphics[clip,width=\linewidth]{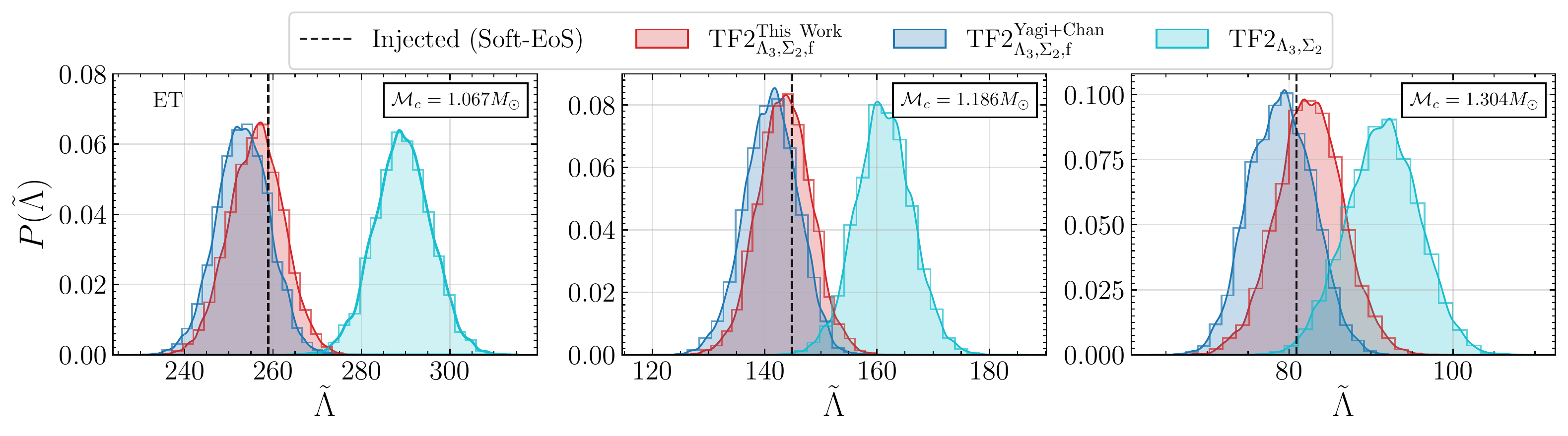}
\label{fig:ltilde_Soft_EoS_ET}
\end{subfigure}

\begin{subfigure}{\linewidth}
\includegraphics[clip,width=\linewidth]{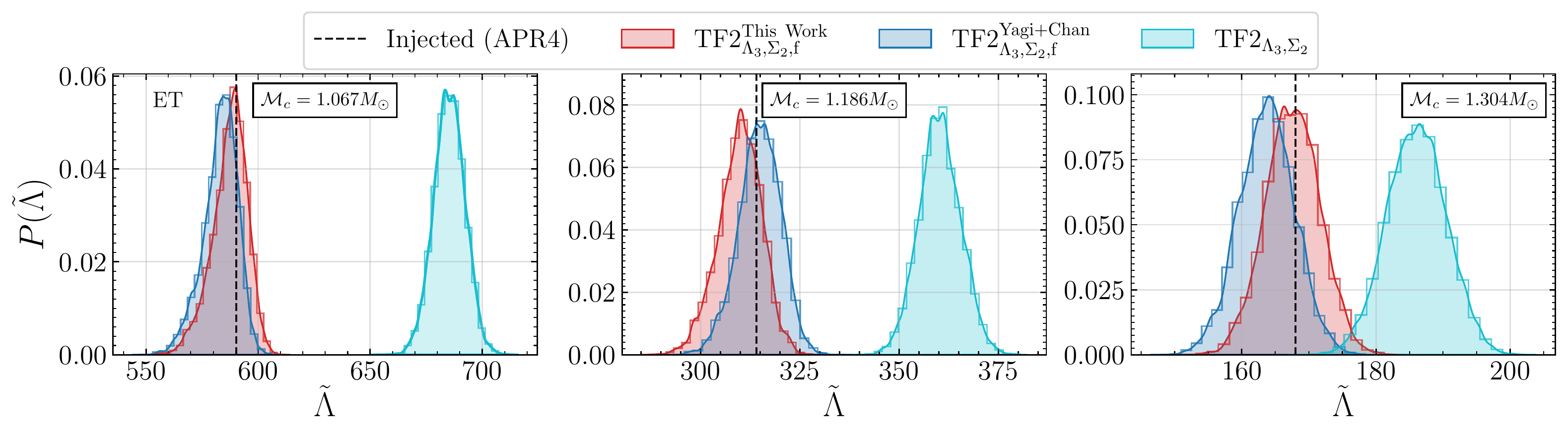}
\label{fig:ltilde_APR4_ET}
\end{subfigure}
 
\begin{subfigure}{\linewidth}
\includegraphics[clip,width=\linewidth]{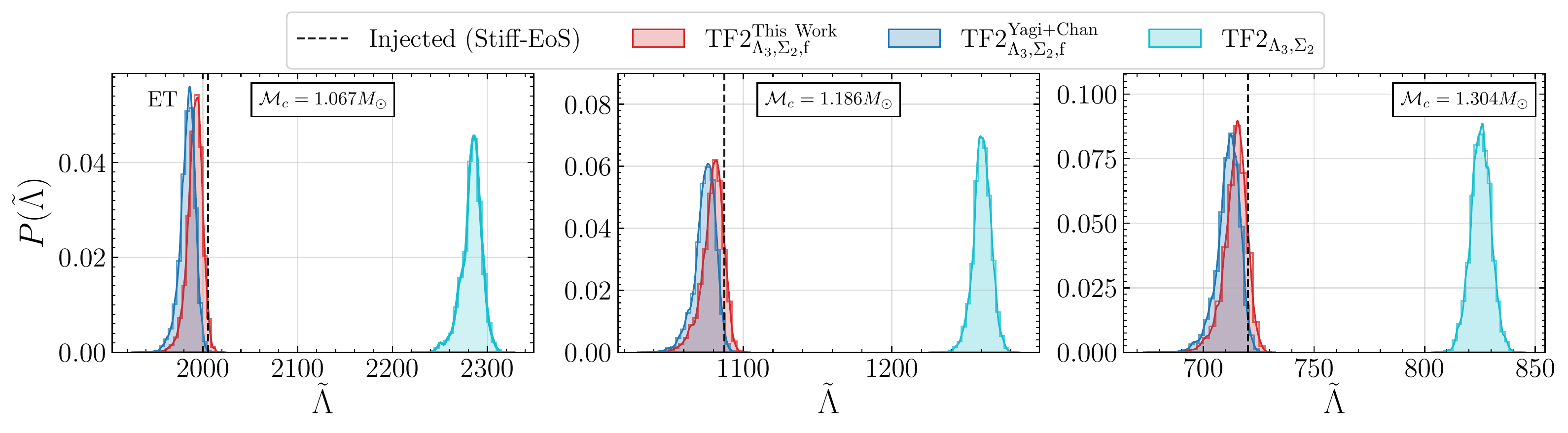}
\label{fig:ltilde_Stiff_EoS_ET}
\end{subfigure}
\caption{Histogram and Posterior distribution of recovered $\tilde{\Lambda}$  for different injection events corresponding to the  Soft-EoS (upper panel), APR4 EoS  (middle panel) and    Stiff-EoS (lower panel). The injected values are shown with black dashed lines. The analysis is done for ET configuration }
  \label{fig:injection_eos_ET}

\end{figure*}

As described in ~\Cref{subsec:injection_studies}, we perform the injection and recovery studies with the third generation Einstein Telescope (ET)  detector with the proposed ET-D sensitivity. We display the distribution of recovered  $\tilde{\Lambda}$ in ET sensitivity in ~\Cref{fig:injection_eos_ET} (similar to ~\Cref{fig:injection_eos})~.

\bibliography{Pradhan}

\end{document}